\documentclass[twocolumn]{aastex62}
\usepackage{natbib}

\usepackage{multirow}
\usepackage{color}
\usepackage{longtable}
\usepackage{soul}

\setstcolor{red}

\shorttitle{Nature of Faint Radio Sources-I Spectral Index and Radio-FIR Correlation}
\shortauthors{Gim et al.}

\begin{document}

\title{Nature of Faint Radio Sources in GOODS-North and GOODS-South Fields -- I. Spectral Index and Radio-FIR Correlation}

\author{Hansung B. Gim}
\affiliation{School of Earth and Space Exploration, Arizona State University, Tempe, AZ 85281, USA}
\affiliation{Department of Astronomy, University of Massachusetts, Amherst, MA 01003, USA}

\author{Min S. Yun}
\affiliation{Department of Astronomy, University of Massachusetts, Amherst, MA 01003, USA}

\author{Frazer N. Owen}
\affiliation{National Radio Astronomy Observatory, P. O. Box O, Socorro, NM 87801, USA}

\author{Emmanuel Momjian}
\affiliation{National Radio Astronomy Observatory, P. O. Box O, Socorro, NM 87801, USA}

\author{Neal A. Miller}
\affiliation{Department of Mathematics and Physical Sciences, Stevenson University, 1525 Greenspring Valley Rd, Stevenson, MD 21153, USA}

\author{Mauro Giavalisco}
\affiliation{Department of Astronomy, University of Massachusetts, Amherst, MA 01003, USA}

\author{Grant Wilson}
\affiliation{Department of Astronomy, University of Massachusetts, Amherst, MA 01003, USA}

\author{James D. Lowenthal}
\affiliation{Department of Astronomy, Smith College, Northampton, MA 01063, USA}

\author{Itziar Aretxaga}
\affiliation{InstitutoNacionaldeAstrof\'{i}sica, \'{O}pticayElectr\'{o}nica,Luis Enrique Erro 1, Sta. Ma. Tonantzintla, 72000, Puebla, Mexico}

\author{David H. Hughes}
\affiliation{InstitutoNacionaldeAstrof\'{i}sica, \'{O}pticayElectr\'{o}nica,Luis Enrique Erro 1, Sta. Ma. Tonantzintla, 72000, Puebla, Mexico}

\author{Glenn E. Morrison}
\affiliation{LBT Observatory, University of Arizona, 933 N. Cherry Avenue, Room 552, Tucson, AZ 85721, USA}
\affiliation{Canada-France-Hawaii Telescope, Kamuela, Hawaii, HI 96743, USA}

\author{Ryohei Kawabe}
\affiliation{National Astronomy Observatory of Japan, Osawa, Mitaka, Tokyo, 181-8588, Japan}

\correspondingauthor{Hansung B. Gim}
\email{Hansung.Gim@asu.edu}

\begin{abstract}
We present the first results from the deep and wide 5~GHz radio observations of the Great Observatories Origins Deep Survey (GOODS)-North ($\sigma = 3.5 \, \mu$Jy beam$^{-1}$, synthesized beam size $\theta = 1.47\arcsec \times1.42\arcsec$, and 52 sources over 109 arcmin$^{2}$) and GOODS-South ($\sigma = 3.0\, \mu$Jy beam$^{-1}$, $\theta = 0.98\arcsec \times0.45\arcsec$, and 88 sources over 190 arcmin$^{2}$) fields using the Karl G. Jansky Very Large Array. We derive radio spectral indices $\alpha$ between 1.4 and 5~GHz using the beam-matched images and show that the overall spectral index distribution is broad even when the measured noise and flux bias are considered. We also find a clustering of faint radio sources around $\alpha=$0.8, but only within $S_{5GHz} < 150\, \mu$Jy. We demonstrate that the correct radio spectral index is important for deriving accurate rest frame radio power and analyzing the radio-FIR correlation, and adopting a single value of $\alpha=$0.8 leads to a significant scatter and a strong bias in the analysis of the radio-FIR correlation, resulting from the broad and asymmetric spectral index distribution. When characterized by specific star formation rates, the starburst population (58\%) dominates the 5~GHz radio source population, and the quiescent galaxy population (30\%) follows a distinct trend in spectral index distribution and the radio-FIR correlation. Lastly, we offer suggestions on sensitivity and angular resolution for future ultra-deep surveys designed to trace the cosmic history of star formation and AGN activity using radio continuum as a probe.

\end{abstract}

\keywords{radio continuum: general ---
radio spectral index, radio-far infrared correlation, star formation: individual(\objectname{GOODS-North},
\objectname{GOODS-South})}

\section{Introduction}
Stellar mass build-up and central massive black-hole growth are two key observational constraints for understanding galaxy evolution in modern astronomy. A significant fraction of these activities are heavily obscured by dust over the cosmic history \citep{lefloch05, caputi07, magnelli11a, whitaker17}, and we need another tracer that can penetrate deep into column densities exceeding $N_{HI} > 10^{24}$ cm$^{-2}$ ($A_{V} \gg 100$). The completion of the NSF's Karl G. Jansky Very Large Array\footnote{The National Radio Astronomy Observatory is a facility of the National Science Foundation operated under cooperative agreement by Associated Universities, Inc.} (VLA) with a more than 100 times larger spectral bandwidth and a new powerful digital correlator translates to more than an order of magnitude improvement in sensitivity to probe star formation and black hole activities at cosmological distances \citep{perley11}. 

The low-frequency ($\nu \lesssim 10$~GHz) radio sky is dominated by synchrotron emission \citep{condon92}, which mainly comes from star-forming galaxies (SFGs) and active galactic nuclei (AGN). In SFGs, synchrotron emission is generated through cooling of cosmic rays accelerated by shocks associated with Type II supernovae. In AGN, synchrotron radiation is produced by relativistic charged particles in radio cores and jets.  Different origins of the observed synchrotron radiation are encoded in radio spectral index $\alpha$, which is defined as $S \propto \nu^{-\alpha}$, where $S$ is the flux density and $\nu$ is the frequency. Star-forming regions are optically thin to synchrotron radiation, which yields a steep, characteristic radio spectral index of $\alpha \approx$ 0.8 \citep{condon92}.  Synchrotron emission in AGN is produced in two different ways. Radio core AGN are optically thick enough to absorb synchrotron emission and re-emit, which makes the slope of the synchrotron radiation flatter (``synchrotron self-absorption''), $\alpha \ll 0.8$ \citep{debruyn76}. In jets, relativistic electrons lose their energy over time while traveling down the length of the jets, and the resulting radio spectral index is steeper (``synchrotron aging''), $\alpha > 0.8$ \citep[e.g.,][]{burch79}. 

Radio spectral indices have been used to study the nature of radio sources. In particular, the emergence of flat spectrum sources in the sub-mJy regime has been reported by several authors \citep[e.g.,][]{donnelly87, prandoni06, randall12}, although others have reported no flattening in the mean spectral index \citep{fomalont91, ibar09}.  Deeper radio observations with $\mu$Jy sensitivity have shown that the fraction of steep spectrum sources increases with decreasing flux density, suggesting the emergence of SFGs at the sub-mJy level \citep{ibar09, huynh15, murphy17}, in agreement with the interpretation of the normalized number counts \citep{owen08, condon12} and the analysis of the polarization \citep{rudnick14}. A radio study of sub-millimeter galaxies (SMG) has showed that their radio spectral index distribution is a skewed Gaussian with a peak near $\alpha\sim0.7$ and a tail towards flatter spectrum \citep{ibar10}. These studies indicate a promising potential for the radio spectral index as a tracer of underlying physical activity in distant galaxies.

We show here that obtaining {\em correct} measurements of radio spectral indices is critically important in calculating the rest-frame radio power and for understanding the cosmic evolution of the faint radio population. Any uncertainty in radio spectral index translates directly to the uncertainty in derived radio power, and this in turn affects the accuracy of the radio-far infrared (FIR) correlation analysis \citep{gim15, delhaize17}. Radio AGNs with jets are often resolved by interferometric observations, and even normal SFGs show spatially resolved structures at arcsecond scales \citep[e.g.,][]{chapman04,barger17}. 

In this paper, we present the analysis of radio spectral indices between 1.4 and 5~GHz derived with matched beams, for a large sample of faint radio sources identified from the deep and wide 5~GHz radio observations on the GOODS-North (GN) and -South (GS) fields. We examine the correlations among radio spectral index, radio-FIR correlation, and star formation properties. We also discuss the limitations of radio observations tracing normal SFGs, the importance of correct derivation of radio spectral index, and the constraints provided by radio spectral index to classifying radio SFGs. Throughout this paper we adopt the cosmological parameters, $H_{0}=$ 67.8 km s$^{-1}$ Mpc$^{-1}$, $\Omega_{m}=$ 0.308, and $\Omega_{\Lambda}=$ 0.692 \citep{pdg18}.

\section{Observations \label{OBS}}

\subsection{Radio Observations \label{RADIO}}
\subsubsection{GOODS-North \label{RADIO_GN}}
Our observations of the GN field were conducted in February and March of 2011, for a total of 22 hours at 5~GHz in the B-configuration of the VLA under the program code {\bf 10C-225}. As summarized in Table~\ref{observations}, we observed two fields with the VLA's Wideband Interferometric Digital Architecture (WIDAR) correlator which was configured to deliver two 128~MHz sub-bands in full polarization. The sub-bands were further split into 64$\times$2~MHz channels each, and centered at 4896 and 5024~MHz, respectively. The correlator integration time was 3 seconds. 

\begin{table}
\begin{center}
\caption{Observation Summary \label{observations}}
\begin{tabular}{ccccc}
\tableline
\tableline
Field & R.A. (J2000) & Dec. (J2000) & Date & Duration \\
\tableline
\multirow{4}{*}{GN} & \multirow{2}{*}{12$^{h}$ 36$^{m}$ 31$^{s}$.3} & \multirow{2}{*}{62$^{\circ}$10\arcmin50.0\arcsec} & 2011 Feb 28 & 5.5 hrs \\
 &  &  & 2011 Mar 10 & 5.5 hrs \\
 & \multirow{2}{*} {12$^{h}$ 37$^{m}$ 07$^{s}$.5} & \multirow{2}{*} {62$^{\circ}$14\arcmin 51.0\arcsec} & 2011 Mar 15 & 5.5 hrs \\
  &  &  & 2011 Mar 20 & 5.5 hrs \\
\hline
\multirow{6}{*}{GS} & 03$^{h}$ 32$^{m}$ 30$^{s}$.00 & -27$^{\circ}$43\arcmin 45.0\arcsec & 2012 Dec 16 & 2.5 hrs \\
  & 03$^{h}$ 32$^{m}$ 13$^{s}$.33 & -27$^{\circ}$45\arcmin 52.5\arcsec & 2012 Dec 23 & 2.5 hrs \\
  & 03$^{h}$ 32$^{m}$ 13$^{s}$.33 & -27$^{\circ}$50\arcmin 07.5\arcsec & 2012 Dec 31 & 2.5 hrs \\
  & 03$^{h}$ 32$^{m}$ 30$^{s}$.00 & -27$^{\circ}$52\arcmin 15.0\arcsec & 2013 Jan 01 & 2.5 hrs \\
  & 03$^{h}$ 32$^{m}$ 46$^{s}$.67 & -27$^{\circ}$50\arcmin 07.5\arcsec & 2013 Jan 05 (1) & 2.5 hrs \\
  & 03$^{h}$ 32$^{m}$ 46$^{s}$.67 & -27$^{\circ}$45\arcmin 52.5\arcsec & 2013 Jan 05 (2) & 2.5 hrs \\
\tableline
\end{tabular}
\end{center}
\end{table}

The calibration and reduction of the VLA data were carried out using the standard data reduction package Astronomical Image Processing System (AIPS). The flux calibrator 3C286 was used for the calibrations of delay, flux density scale, and polarization while the gain calibrator J1400+6210 was used for the bandpass and gain calibration. The radio quasar J1400+6210 is bright enough (1.72~Jy at 5~GHz) to be used for the bandpass calibration. 

\begin{figure*}
\includegraphics[angle=0,scale=0.68]{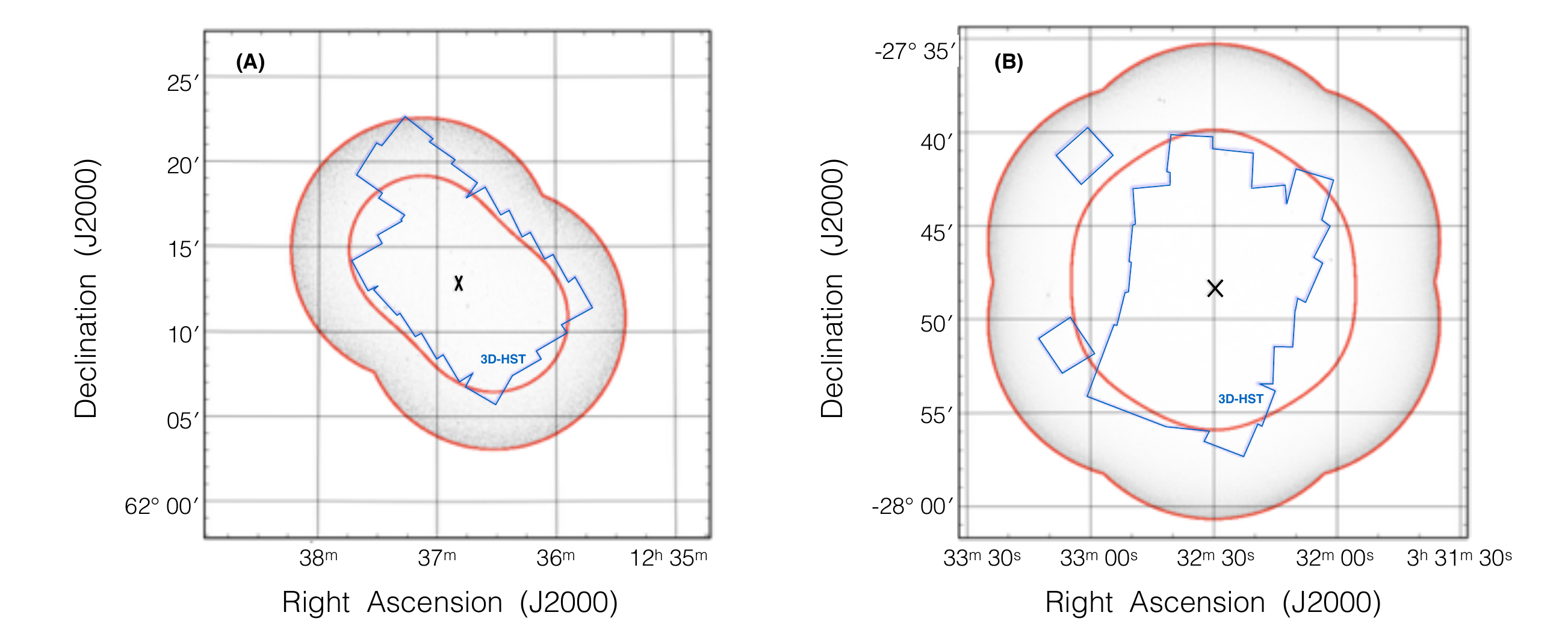}
\caption{The VLA 5 GHz mosaic images for the GN (panel A) and the GS (panel B) fields.  The inner red contours trace the boundary where the primary beam correction increases the effective noise to twice that in the image center, and this also marks the survey area where the source catalogs are derived.  The outer red contours mark the survey areas where the primary beam response is 7\% of the mosaic center.  The centers of each mosaic are marked with a cross (`$\times$'), and the 3D-HST coverages are outlined in blue polygons. \label{coverage}}
\end{figure*}

Imaging of the visibility data was performed using the Common Astronomy Software Applications \citep[CASA,][]{mcmullin07}. The wide field imaging of each field was carried out using nine facets, each with 4096$\times$4096 pixels with a cell size of 0.35\arcsec, down to 3$\sigma$. The Clark point spread function (PSF) model is adopted, and the Briggs function is used to weight the data with a robust value of $R=1$. The Briggs weighting function is intermediate between natural (lowest noise, poorest resolution) and uniform (highest noise, best resolution) weighting functions, and the robust factor of $R=1$ gives an optimal compromise between sensitivity and resolution. The final mosaic and sensitivity images incorporating the primary beam correction are produced using the AIPS tasks LTESS and STESS, respectively. The final mosaic image has a size of 5120$\times$5120 pixels, centered at [12$^{h}$ 36$^{m}$ 49$^{s}$.4, 62$^{\circ}$ 12\arcmin\ 50.5\arcsec] (J2000), with a synthesized beam of 1.47\arcsec$\times$1.42\arcsec. The effective central frequency of the image is 4.959~GHz (hereafter 5~GHz) with a total bandwidth of 240 MHz. The final noise is $\sigma=3.5$ $\mu$Jy beam$^{-1}$ in the image center.  The survey coverage map for the GN field is shown in panel (A) of Figure~\ref{coverage}. 

\subsubsection{GOODS-South \label{RADIO_GS}}
The GS field was observed at 5~GHz for a total of 15 hours in the A-configuration of the VLA under the project code of {\bf 12B-274}. The coordinates of the six pointing centers and observation dates are listed in Table~\ref{observations}. The WIDAR correlator was configured to deliver sixteen 128~MHz sub-bands, each with 64$\times$2~MHz channels and full polarization products. The frequency span was from 4488 to 6512~MHz. Correlator integration time was 1 second to minimize the time smearing effect.  The observations were executed in six different sessions, each with 2.5 hrs long. 

Data reduction and imaging were performed using CASA. The flux density scale calibrator 3C48 was used for the calibrations of delay, flux scale, and polarization, while the gain calibrator J0240$-$2309 (2.33Jy at 5~GHz) was used for calibrations of bandpass, phase, and delay. Severe radio-frequency interference (RFI) dominated the last four SPWs (12 to 15), and they are excluded in the analysis. Self-calibration was carried out successfully to improve the overall dynamic range of the image using bright sources ($>$ 1~mJy) in each field.

Initial imaging was done in CASA for each field and each SPW exploiting the wide-field imaging with 36 facets that are each 10240$\times$10240 pixels in size and using a cell size of 0.1\arcsec, down to 3$\sigma$. The Clark PSF and the Briggs weighting function with a robust value of $R=0.8$ are adopted for imaging. The synthesized beam depends on the frequency, and all images are convolved to match the largest beam at the lowest frequency SPW before the final mosaic image is constructed. Using the weights of $w_{i}=(beam\; area)_{new}/(beam\; area)_{old}$, all images were convolved to have beam sizes of 0.98\arcsec$\times$0.45\arcsec. The mosaic image of each SPW is produced first using the AIPS tasks LTESS and STESS with primary beam correction. The final band-merged image is produced by averaging the SPW mosaic images using the $1/\sigma^{2}$ weight, where $\sigma$ is an RMS noise of each mosaic. The final band-merged mosaic image is 16384$\times$16384 pixels in size with the central frequency of 5.245~GHz (hereafter 5~GHz) and a total bandwidth of 1.486~GHz.  The RMS noise in the center of the mosaic is  $\sigma = 3.0\, \mu$Jy beam$^{-1}$, and the coverage map centered on [03$^{h}$ 32$^{m}$ 30$^{s}$.0, $-$27$^{\circ}$ 48\arcmin~00\arcsec] is shown in panel (B) of Figure~\ref{coverage}.

\subsubsection{Source Catalogs \label{CATALOG}}
The 5~GHz sources are extracted from primary-beam corrected images using the AIPS task SAD. Since radio-frequency interference is time-dependent and the primary beam response is not uniform, the final noise distribution is not uniform or symmetric across the mosaic.  Therefore, we limit the source search for generating the catalogs to the central regions with up to twice the RMS noise in primary-beam corrected maps, i.e., $7\, \mu$Jy beam$^{-1}$ for the GN field and $6\, \mu$Jy beam$^{-1}$ for the GS fields as shown with inner red contours in Figure~\ref{coverage}. We also minimized the impact of the effective frequency shift to lower frequency toward the edge of the final image. Since the coverage of the image is different at each SPW due to the frequency-dependent primary beam correction, the effective frequency moves to the lower frequency toward the edge of the frequency-stacked image. We have created a matching sensitivity map to track the frequency-dependent effects in the final mosaic. We also limited our catalog to the more central region reasonably far away from the edges. Sources detected with a peak signal-to-noise ratio (SNR) $>$~5 are selected for the final catalogs, and the measured flux densities are corrected for bandwidth smearing by setting the AIPS adverb BWSMEAR as the fraction of channel width with respect to the central frequency in the SAD. However, the time averaging effect is not taken into account since its impact on the flux density is small ($<$0.1\%) enough to be neglected within our catalog regions \citep{bridle99}. The final catalogs include 52 \& 88 sources in the GN \& GS fields covering 109 \& 190  arcmin$^{2}$ areas, respectively. These catalogs are shown in Appendix~\ref{CAT}. 

\subsubsection{Comparisons with previous results \label{PREVOBS}}
There are recent radio continuum observations of both GN \& GS fields with comparable or higher sensitivity and at a higher angular resolution, and they offer an interesting and complementary view on the nature of the faint radio source population.   \citet{guidetti17} have studied the GN field at 5.5~GHz with an RMS noise of $3\, \mu$Jy beam$^{-1}$ and a synthesized beam size of 0.5\arcsec, and they reported a total of 94 sources ($\ge5\sigma$) over their 154 arcmin$^{2}$ survey area.  This is about 80\% larger number of sources over a 50\% larger area with a similar flux density sensitivity compared to our survey. At least part of this difference must be due to their 3 times smaller beam (9 times worse surface brightness sensitivity), which can fragment some of the resolved star-forming galaxies and jet sources into multiple components. \citet{guidetti17} also suggested this surface brightness sensitivity effect as the root cause for their unexpectedly large (80\%) AGN fraction.

Earlier surveys of the GS field by \citet{kellermann08} at 4.9 GHz using the VLA and by \citet{huynh15} 5.5 GHz using the Australia Telescope Compact Array were both about a factor of 2 shallower in sensitivity ($\sigma\approx8\, \mu$Jy) and 2-3 times lower in angular resolution ($\theta\approx 4\arcsec$) compared to our survey.  \citet{huynh15} reported finding 212 source components over their 0.34 deg$^2$ survey area down to a flux density of $\sim50$ $\mu$Jy ($\ge5\sigma$). \citet{kellermann08} did not report the source count in their 4.9 GHz VLA survey, but \citet{huynh15} reported their data to be consistent because of their similar resolution and sensitivity.  The 5 GHz source density derived from these surveys with $\sim3$ times shallower depth is 2.6 times lower than our survey.  

More recently, \citet{rujopakarn16} have observed the Hubble Ultra Deep Field (HUDF) within the GOODS-South at 6~GHz with an RMS noise of $0.32\, \mu$Jy beam$^{-1}$ at an angular resolution of 0.61\arcsec$\times$0.31\arcsec.  A direct comparison of the source density is difficult in this case because these authors report two source counts that are not fully reflective of the true source density: (1) a total of 68 ``bright" ($\ge8\sigma$) sources within the 61 arcmin$^2$ survey region extending beyond the primary beam; and (2) a total of 11 sources detected at $\ge5\sigma$ among the 13 sources detected by ALMA inside the 40.7 arcmin$^{2}$ ALMA survey area. The former number offers a more useful comparison, and corresponds to about 2.5 times higher source density at 6-8 times better sensitivity compared with our survey.  The latter number is strictly a lower limit since it includes only ALMA-detected sources at $z=1-3$.  The resulting source density is only 60\% of the source density we derive, despite their 10 times better flux density sensitivity.  

In summary, the source density we derive is consistent with those of the past surveys.  A striking trend seen is that the derived source density increases relatively slowly with improved sensitivity.  There are potentially important systematic differences in how the catalogs are generated, and these source counts are not corrected for completeness in a consistent way.  Nevertheless, the rise in source density with improving depth of the survey is far flatter than the Euclidean case. Along with the improving sensitivity, subsequent observations have also employed higher angular resolution, and this might play an important role in the derived source statistics, as discussed further below in \S~\ref{RESOLUTION}.  This also serves as one of our motivations for using beam-matched data for our spectral index analysis (see \S~\ref{ALPHA}).

\subsection{Multi-wavelength data \label{MULTI}}
\subsubsection{VLA 1.4~GHz \label{DATA14}}
1.4~GHz data are needed to calculate the radio spectral index with our 5~GHz data. For the GN field, we use the deep 1.525~GHz (hereafter 1.5~GHz) imaging data obtained by \citet{owen18} with RMS noise of 2.2$\, \mu$Jy beam$^{-1}$ and an angular resolution of 1.6\arcsec$\times$1.6\arcsec (FWHM). \citet{owen18} have used different beam sizes (2\arcsec, 3\arcsec, 6\arcsec, and 12\arcsec) to measure the flux densities of extended sources because those sources were resolved out with the original beam size, which resulted in the prevention of the loss of flux densities. All of our 5~GHz sources have a matching counterpart in the 1.5~GHz source catalog. 

For the GS field, we use the 1.4~GHz VLA data by \citet{miller13}, which has RMS noise of $\sim$6 $\mu$Jy beam$^{-1}$ at the image center with a beam size of 2.8\arcsec$\times$1.6\arcsec. Since the beam area of these 1.4~GHz data is about ten times larger than our 5~GHz data and the depth of the 1.4 GHz data is significantly shallower than in the GN field, matching the counterparts to the 5~GHz sources is more complicated. We convolve the 5~GHz images for each field and SPW to yield a beam size of 2.8\arcsec$\times$1.6\arcsec~using the AIPS task CONVL, and the final mosaic is produced by summing over all pointings and SPW using the AIPS tasks LTESS and STESS.\footnote{Since the final radio image is a combination of cleaned components with flux density scaled by clean beam and residuals with flux densities weighted by dirty beam, the convolution of the radio image with the clean beam includes the convolution of the residuals scaled by the dirty beam in addition to the convolution of the clean components scaled by the clean beam. The former contributes on the uncertainty of the convolved images, but it is not easy to estimate its contributions because it involves many parameters such as clean thresholds, PSF shape, and the convolution kernel size. This is a subtle but  notable systematic effect that we have decided to ignore for the moment.} The RMS noise of the convolved 5~GHz image is slightly higher, 6.4 $\mu$Jy beam$^{-1}$. We generated the 3$\sigma$ catalog from the convolved image using the AIPS task SAD. For the 38 sources that were not found in this 3$\sigma$ catalog due to increased noise and low completeness at low SNR, we manually performed aperture photometry on the convolved image centered on the source coordinates from the original, full resolution image. A total of 83 sources are identified in the final convolved 5~GHz mosaic image with a beam size of 2.8\arcsec$\times$1.6\arcsec, as eight of the sources in the original catalog are now blended into three sources. Matching the 1.4~GHz catalog with this beam-matched 5 GHz data yields 64 counterparts among the 83 sources. A total of 19 sources lack a 1.4~GHz counterpart because the 1.4 GHz data are too shallow (5$\sigma$ $\geq$ 30~$\mu$Jy beam$^{-1}$ at the image center) to detect 5~GHz sources with a flat or inverted spectrum which is a characteristic of some of the radio AGNs (see the panels (D) and (E) of Figure~\ref{alpha_flux}). Throughout this paper, we analyze only the GS sources that have a unique 1.4~GHz counterpart to avoid the uncertainty introduced by the upper limits.

\subsubsection{Chandra X-ray Observatory \label{XRAY}}
We use X-ray data taken from the {\em Chandra X-ray Observatory} survey with full band (0.5-7 keV), soft band (0.5-2 keV), and hard band (2-7 keV) catalogs. We make use of 2~Ms observations for the GN field \citep{xue16} and 7~Ms observations for the GS field \citep{luo17}. The limiting fluxes for the GN field are $3.5 \times 10^{-17}$, $1.2 \times 10^{-17}$, $5.9 \times 10^{-17}$ erg cm$^{-2}$ s$^{-1}$ at full band, soft band, and hard band respectively. For the GS field, the limiting fluxes are $1.9 \times 10^{-17}$ at full band, $6.4 \times 10^{-18}$ at soft band, and $2.7 \times 10^{-17}$ erg cm$^{-2}$ s$^{-1}$ at hard band. To calculate the X-ray luminosity, we assume a photon index of $\Gamma=1.8$ for X-ray detected radio sources \citep{tozzi06} but $\Gamma=1.4$ for X-ray undetected radio sources \citep{luo17}. The full band X-ray luminosity at [0.5-7 keV] is converted to the luminosity at [0.5-8 keV] using the relation of $L_{[0.5-8 keV]} = 1.066 \times L_{[0.5-7keV]}$ for the assumed $\Gamma=1.8$ \citep{xue16}.

\subsubsection{Spitzer Space Telescope \label{SPITZER}}
We exploit publicly released $Spitzer$ $Space$ $Telescope$ ($Spitzer$) IRAC catalogs of the GN \citep{wang10} and GS \citep{damen11} fields. The GN field IRAC catalog has a sensitivity (1$\sigma$) of 0.15$\mu$Jy at 3.6$\mu$m, while the GS field IRAC catalog by the $Spitzer$ IRAC/MUSYC Public Legacy Survey in the Extended Chandra Deep Field-South (ECDFS) has a sensitivity (1$\sigma$) of 0.22$\mu$Jy at 3.6$\mu$m.  We make use of the high angular resolutions of our radio observations to find counterparts within the beam sizes, i.e., 1.47\arcsec\ for the GN and 0.98\arcsec\ for the GS fields.

\subsubsection{Herschel Space Observatory \label{HERSCHEL}}
The comparison FIR data are constructed using the public archival data for the Photodetector Array Camera and Spectrometer (PACS) and the Spectral and Photometric Imaging REceiver (SPIRE) of the $Herschel$ $Space$ $Observatory$\footnote{Herschel is an ESA space observatory with science instruments provided by European-led Principal Investigator consortia and with important participation from NASA.}. The PACS photometry data at 70, 100, and 160 $\mu$m are taken from the combination of PACS Evolutionary Probe \citep[PEP]{lutz11} and GOODS-Herschel \citep{elbaz11} programs described by \citet{magnelli13}. The SPIRE 250, 350, and 500 $\mu$m photometry data are taken from the Herschel Multi-tiered Extragalactic Survey (HerMES) DR 3 and 4 \citep{roseboom10, magnelli11b, roseboom12}. We adopt the catalogs extracted using the $Spitzer$ MIPS 24 $\mu$m position priors for the PACS bands by the GOODS-Herschel collaboration\footnote{Data are available at http://www.mpe.mpg.de/ir/Research/PEP/DR1}. As for the SPIRE bands, we used the catalogs extracted at the SPIRE 250 $\mu$m source positions (HerMES DR4)\footnote{Data are available at http://hedam.lam.fr/HerMES/index/dr4}.

To identify FIR counterparts to the radio sources, we apply the likelihood ratio technique \citep{sutherland92}. The search radius adopted is three times the combined positional uncertainties of the radio and Herschel sources.  Sources with the reliability of $Rel_{i}>$0.8\footnote{Reliability is defined as $Rel_{i} = LR_{i}/(\Sigma LR_{i} + (1-q_{0}))$ for the likelihood ratio $LR_{i}$ and the fraction of true counterparts above the detection limit, $q_{0}$} are accepted as formal counterparts. We consider an FIR source to be the counterpart to a radio source if it is detected in at least one band in both PACS and SPIRE, with a SNR$>4$ in at least one band.

We have compiled the observed 24, 100, 160, 250, 350, and 500 $\mu$m band fluxes of 40 GN and 44 GS sources. The best-fit FIR SED models are identified using a widely used SED fitting code $Le$ $Phare$\footnote{$Le$ $Phare$ is available at http://www.cfht.hawaii.edu/ $\sim$arnouts/lephare.html} \citep{arnouts99, ilbert06} with various SED templates for SFGs \citep{chary01,dale01,lagache03} and QSOs \citep{polletta07}.  This analysis yielded a good SED model for 39 GN  and 42 GS sources. For the radio sources undetected at FIR or with a poor-fit SED, we calculate IR luminosity with 4$\sigma$ flux limits adopting the average $z=1$ SFG SED template \citep{kirkpatrick12}.

\subsubsection{Spectroscopic redshifts \label{SPECZ}}
Spectroscopic redshifts are compiled from the published surveys: GN \citep{cowie04a,donley07,barger08,wirth15} and GS \citep{szokoly04,zheng04,mignoli05,ravikumar07,vanzella08,popesso09,straughn09,balestra10,silverman10,cooper12,kurk13,lefevre14,skelton14,morris15}, respectively. From these compilations, we have 45 (out of 52) sources with spectroscopic redshifts for the GN and 55 (out of 64) for the GS field. In particular, all 55 GS sources with a spectroscopic redshift are in the subsample of 64 sources with both 1.4 GHz and 5 GHz photometry used for the spectral index analysis. 

Even though reliable photometric redshifts from well-sampled photometry data exist in both fields, we limit our analysis to only those with a spectroscopic redshift because errors in redshift translate directly to a large scatter and systematic biases in the derived quantities such as the rest frame radio power, radio-FIR correlation, and star formation rate (SFR).  A detailed evaluation of the accuracy of the existing photometric redshifts and a quantitative analysis on the magnitude of error introduced by using photometric redshifts using this spectroscopic subsample are presented in Appendix~\ref{ZPHOT}. 
Adding those sources with only photometric redshifts to our statistical analysis can in principle expand our sample by up to 16\%, but we have elected to remove this major source of scatter in our statistical analyses presented here for now.

\subsubsection{3D-HST \label{3DHST}}
We adopt physical parameters such as stellar mass, SFR, and effective radius for our 5~GHz sources that also appear in the 3D-HST \footnote{This work is based on observations taken by the 3D-HST Treasury Program (GO 12177 and 12328) with the NASA/ESA HST, which is operated by the Association of Universities for Research in Astronomy, Inc., under NASA contract NAS5-26555.} \citep{brammer12,skelton14,momcheva16} database.  Stellar mass is estimated by the FAST code \citep{kriek09} with the \citet{chabrier03} initial mass function, and the \citet{bc03} stellar population synthesis library \citep{skelton14}. The SFR is computed through the conversion of UV+IR luminosity, where UV luminosity is derived from the rest-frame luminosity at 2800\AA, and IR [8-1000$\mu$m] luminosity is derived from $Spitzer$ MIPS 24$\mu$m flux density by assuming the log average of \citet{dh02} templates \citep[see][]{whitaker14}. Effective radius (R$_{eff}$) is  the semi-major axis of the ellipse containing one half of the total flux of the best S\'{e}rsic model given by GALFIT \citep{vanderwel12}. 

\begin{figure}
\center
\includegraphics[angle=0,scale=.75]{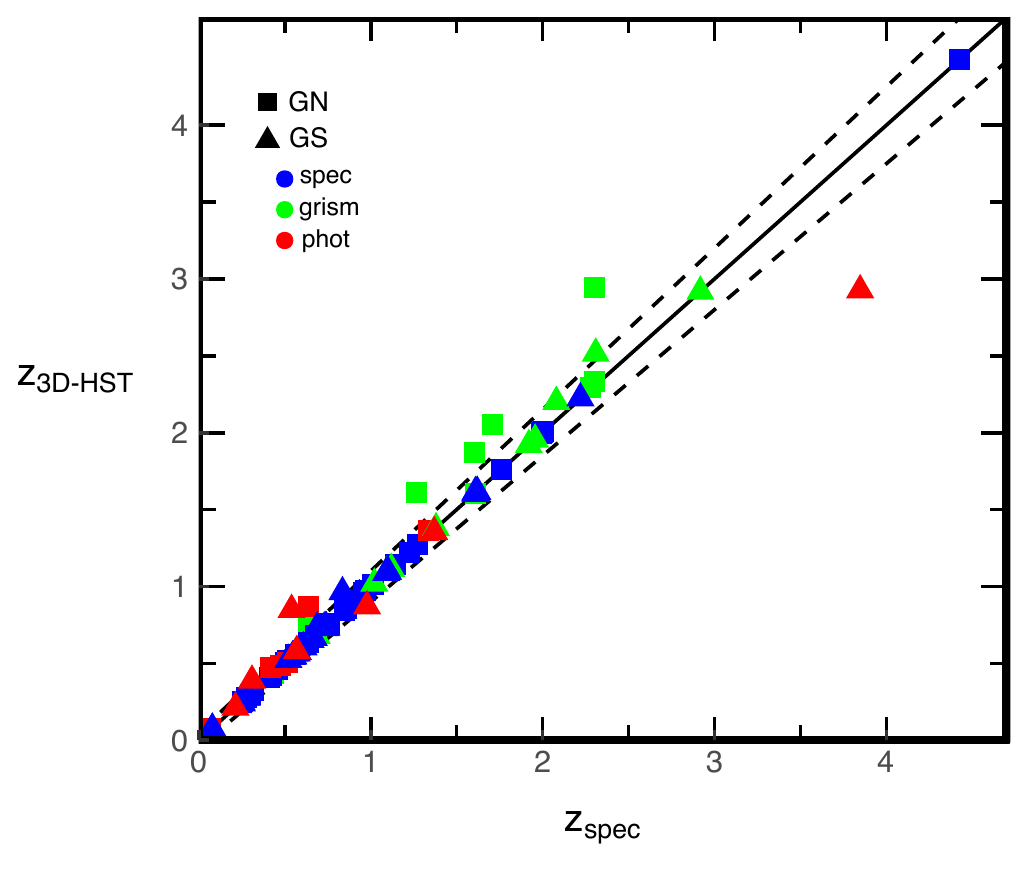}
\caption{Comparison of our spectroscopic redshifts and the best redshifts in 3D-HST for the GN (square) and the GS (triangle) fields. The color represents the type of redshift measurements, e.g. spectroscopy (blue), grism (green), and photometry (red). The solid line is the one-to-one line and dashed lines show the selection limits of $\pm 0.05$ in $|z_{spec}-z_{best,3D-HST}|/(1+z_{spec}) < 0.05$. \label{specz_3dhst}}
\end{figure}

The spectroscopic redshifts given in the 3D-HST database are not as complete as our compilation, and we have to match our spectroscopic redshifts with the best redshifts in the 3D-HST database, which ranks them by spectroscopic, grism, or photometric redshift.  A comparison of the best 3D-HST redshifts with our spectroscopic redshifts is shown in Figure~\ref{specz_3dhst}. We choose the 3D-HST counterparts with best redshifts satisfying $|z_{spec}-z_{best,3D-HST}|/(1+z_{spec}) < 0.05$, which is shown with dashed lines in Figure~\ref{specz_3dhst}.  Spectroscopic redshifts of the best redshifts in 3D-HST are mostly the same as ours while there are some small to significant offsets in grism and photometric redshifts. Through matching the redshifts, we have 3D-HST counterparts for 39 GN and 45 GS radio sources.

\begin{figure*}
\center
\includegraphics[angle=0,scale=.75]{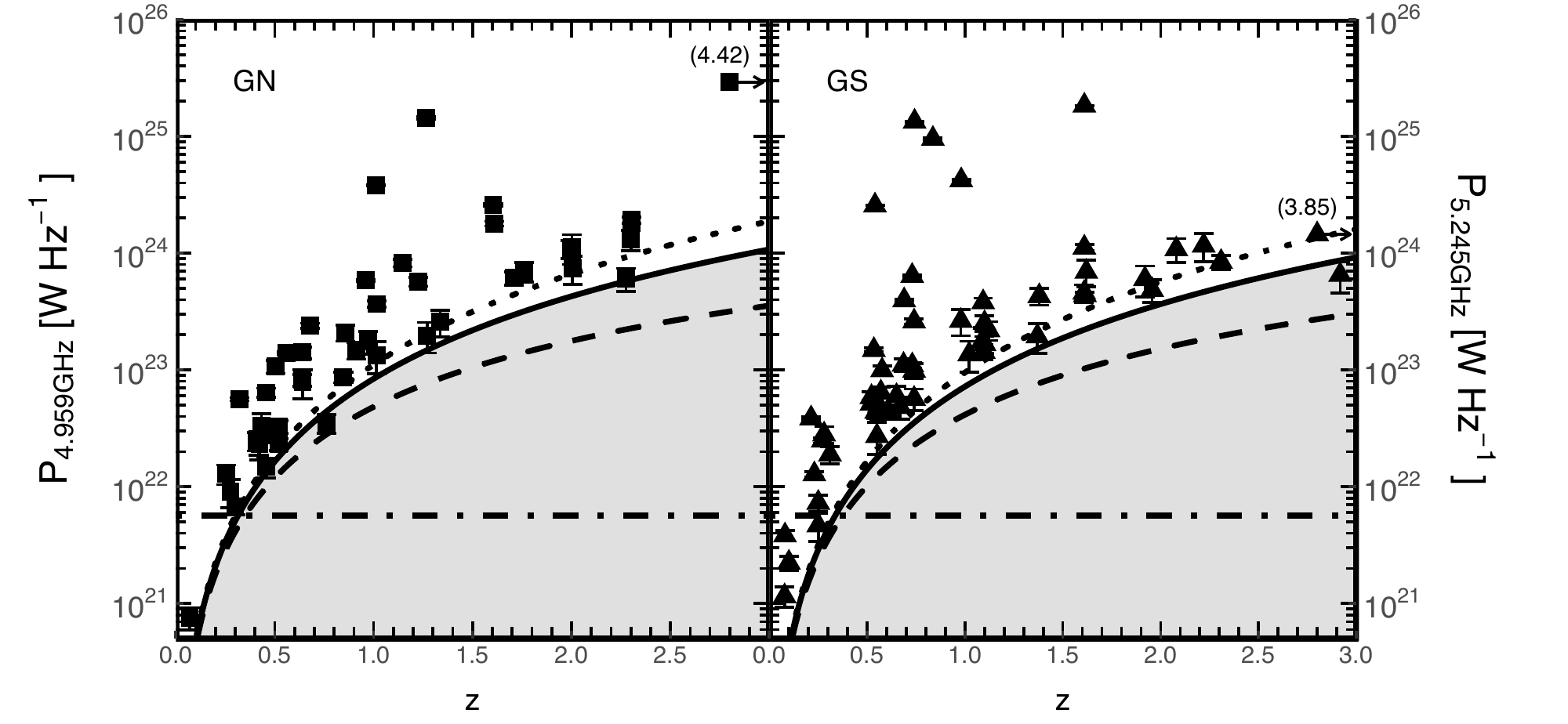}
\caption{Rest-frame 5~GHz radio power as a function of redshift. Only the radio sources with a spectroscopic redshift are shown.  The $5\sigma$ detection limits are shown for radio spectral index of $\alpha=$0 (dashed line), 0.8 (solid line), and 1.2 (dotted line), where $S \propto \nu^{-\alpha}$. The horizontal dot-dashed line at $P_{5GHz}=5.7 \times 10^{21}$W Hz$^{-1}$ corresponds to a luminous infrared galaxy (LIRG) with a $SFR=10 M_{\odot}$ yr$^{-1}$.  Sources with redshift beyond 3 are marked with arrows and their redshifts are written inside the parentheses. \label{lumz}}
\end{figure*}

\section{Selection Function and Rest Frame 5 GHz Radio Power \label{P5GHz}}

In Figure~\ref{lumz}, we show the selection function of our radio sources with rest-frame 5~GHz radio power as a function of redshift. The rest-frame radio power is calculated using the measured spectral index as 
\begin{equation}
 P_{5GHz}=4 \pi d_{L}^{2} S_{5GHz} (1+z)^{\alpha-1} [{\rm W\, Hz}^{-1}], 
\end{equation}
where $d_L$ is the luminosity distance, $S_{5GHz}$ is the measured 5 GHz flux density of the original map, and $\alpha$ is the measured radio spectral index between 1.4 GHz and 5 GHz using the convolved map (see \S~\ref{ALPHA}). The strong positive k-correction associated with radio sources translates to a significant selection bias in favor of flat spectrum sources ($\alpha=0$, dashed line) with lower intrinsic radio power, but such flat spectrum sources are rare in our sample, as shown in this plot (also see Fig.~\ref{alpha_flux}).
The selection functions of the two fields are similar with comparable mean and median values of 5 GHz radio power and redshifts, and a joint analysis of the combined sample is reasonable as long as the slight difference in the catalog depth is properly taken into account. 

The majority of the detected sources have rest frame 5 GHz radio power between $10^{22}$ and $10^{24}$ W Hz$^{-1}$, which is the range of radio power associated with intense starburst systems (LIRGs, ULIRGs) and Seyfert nuclei in the local universe.  However, gas content and SFR of star forming main sequence (MS) galaxies are known to increase rapidly with increasing redshift by an order of magnitude to $z\ge1$ \citep[e.g.,][]{speagle14,scoville17}, and a large fraction of these galaxies at higher redshifts are likely powered by star formation, as discussed below.
Only four sources (two in each field) have a radio power high enough to be classified as ``radio-loud" with $P_{5GHz}\ge 10^{25}$ W Hz$^{-1}$ \citep{miller90}.

\section{Radio Spectral Index \label{ALPHA}}

Radio spectral index $\alpha$ is a measure of the shape of a radio spectrum characterized as a power-law, $S \sim \nu^{-\alpha}$.  We compute the spectral index between 1.4 and 5~GHz using the flux densities derived from the 5 GHz images beam-matched to the 1.4~GHz images as described in Section \ref{DATA14}. In principle, the radio spectral index can be estimated using only the 5 GHz data with its wide bandwidth of 1.5 GHz through the multi-frequency synthesis. The algorithm that can produce in-band spectral index calculation for mosaic observations was not available in CASA when the data were being analyzed. The significant changes in the size of both the primary beam and the synthesized beam across the bandwidth make this in-band spectral index calculation challenging, especially away from the pointing center. These difficulties result in the errors of the in-band spectral index which are not competitive with those using the full 1.4-5 GHz spectral baseline. It is empirically shown that the majority of radio sources in a wide range of redshifts show radio spectra that are fit well with a simple power-law \citep[e.g.,][]{klamer06}. In the frequency range between 1.4 and 5 GHz, the contribution by free-free emission is generally negligibly small  \citep{condon92}.  

\begin{figure*}
\includegraphics[angle=0,scale=0.75]{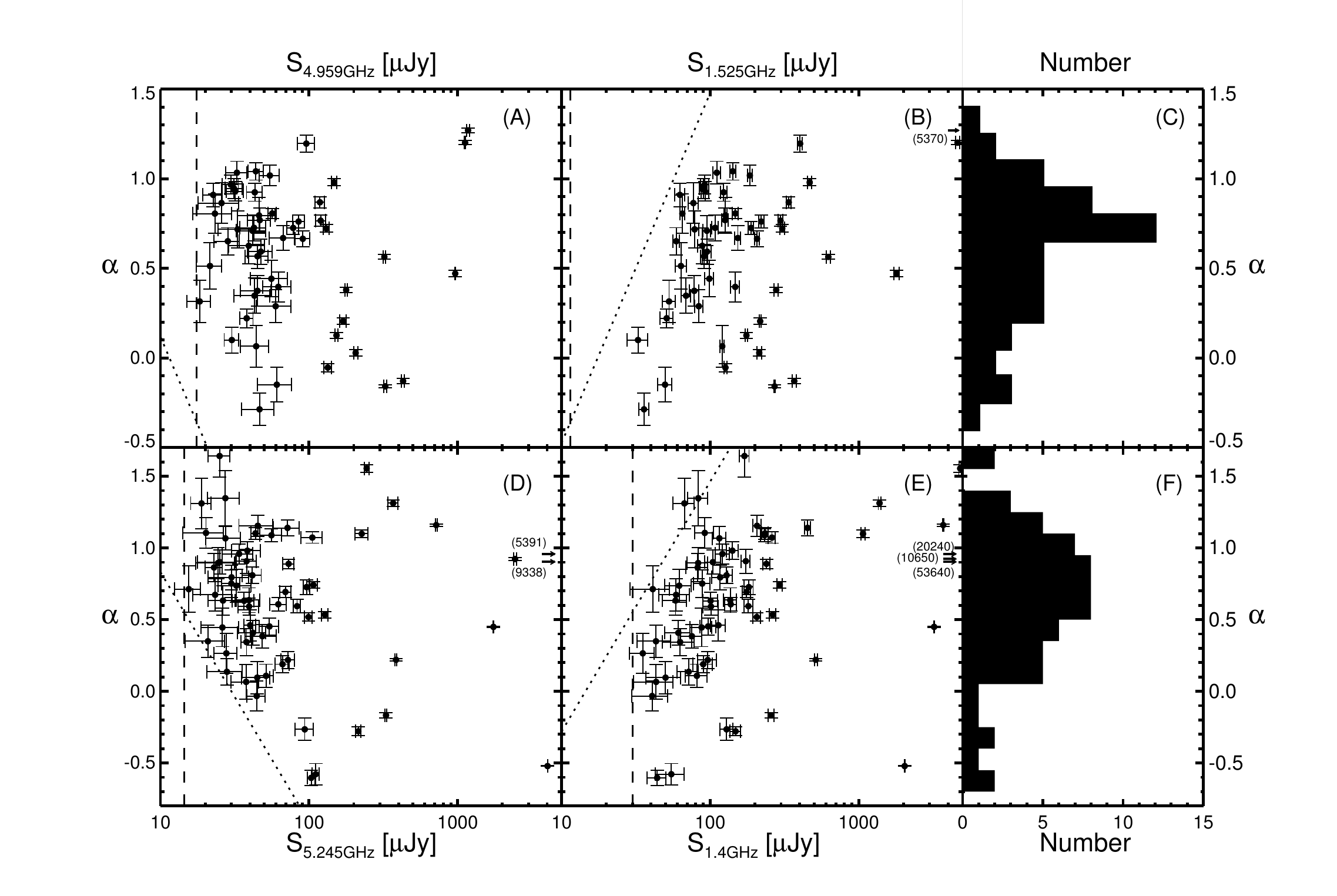}
\caption{Radio spectral index as a function of flux density. The radio spectral indices of our 5~GHz radio sources are plotted as functions of flux densities at 5.0~GHz (panel A), 1.5~GHz (panel B) for the GN field, and 5.2~GHz (panel D) and 1.4~GHz (panel E) for the GS field with 5$\sigma$ flux limits of each survey (dashed line), i.e. 17.5$\mu$Jy (A), 11.5$\mu$Jy (B), 15$\mu$Jy (D), and 30$\mu$Jy (E). The sources located outside the panels are marked with arrows, and their flux densities are given inside the parentheses. The dotted line in each panel represents the sensitivity of radio spectral index limited by the survey limit of adjacent survey.  Histograms of SI distribution for the GN and GS fields are shown in panels (C) and (F), respectively.  \label{alpha_flux}}
\end{figure*}

The distributions of radio spectral index as a function of flux density are shown in Figure~\ref{alpha_flux}. Panels (A), (B), and (C) are for the GN field while the panels (D), (E), and (F) are for the GS field. Since the sensitivity for radio spectral index (dotted line) depends on the flux density limit of the second band (dashed lines), it is not uniform as a function of flux density, and this is a common but important feature for all flux-limited surveys.  Specifically, this non-uniform completeness limits our study to a narrower range of radio spectral indices at fainter flux densities. For our 5~GHz selected sample analyzed here, the depth of the existing 1.4~GHz data restricts the observable range of radio spectral index.  We can see this effect clearly in panel (D), where the range of the radio spectral indices is limited to $\alpha > 0$ even at $S_{5GHz}=30\, \mu$Jy (10$\sigma$), and this can potentially lead to missing sources with inverted spectra at flux densities of $S_{5GHz} < 30\, \mu$Jy.  In practice, however, few inverted spectrum sources with $S_{5GHz} < 35\, \mu$Jy ($10\sigma$) are found in the GN field (panel A), and the actual impact of this potential bias may be limited. 

The uncertainties in the derived radio spectral indices are mainly attributed to the larger uncertainties of flux densities at 5~GHz for the GN field and flux densities at 1.4~GHz for the GS field. The radio spectral index distribution in the GS field is broader and smoother than that in the GN field, and this can be attributed to the shallow depth of the 1.4~GHz data and the noisier 5~GHz photometry as a result of the convolution with a larger Gaussian kernel. Another source of the uncertainty is the wide bandwidth of the VLA. The effective frequency of each flux density measurement depends on the bandwidth and the spectral shape of the source, and this could lead to a significant offset of the effective frequency from the instrumental frequency.  For the steepest spectrum source with $\alpha=$1.64 in the GS field, we estimate that this effect can lead to a maximum frequency offset of 0.1 GHz and a maximum deviation of 0.02 in the derived radio spectral index. Thus, we conclude that this effect has only a minor impact on our radio spectral index calculation. When these systematic effects are taken into account, the distributions of radio spectral indices in these two fields are consistent with each other. 

Panel (A) in Figure~\ref{alpha_flux} shows a clustering of radio sources at $\alpha \sim$0.75 and $S_{5GHz} \le 150\, \mu$Jy, leading to a prominent peak in the histogram in panel (C).  The peak of the radio spectral index histogram for the GS field (panel F) occurs at the same $\alpha$ value, but the clustering is not as pronounced, possibly diluted and broadened by the larger uncertainties in the measured radio spectral indices (see panels B \& E). This peak in the $\alpha$ of steep spectrum radio sources at $S_{5GHz} \le 150\, \mu$Jy has not been reported by earlier studies \citep[e.g.,][]{donnelly87, fomalont91}, but their small sample size (30 in \citealt{donnelly87} and 41 in \citealt{fomalont91}) likely contributed to their poor statistics. 
A more recent study of a larger sample by \citet{huynh15}, who measured radio spectral indices of 5.5~GHz selected sources above $S_{5.5GHz} \gtrsim 50\mu$Jy in the Extended Chandra Deep Field-South (ECDFS) using the 1.4~GHz catalog of \citet{miller13}, did report a spectral index distribution with a clear peak near $\alpha\sim0.7$, as long expected of the star forming galaxy population (see the discussion below).
We note that \citet{huynh15} computed their radio spectral index without matching the beam sizes (about a factor of 2.2 in diameter), and this might be a source of an important systematic error -- see further discussions in \S~\ref{DISC_SI}.  

A natural explanation for the peak near $\alpha \sim 0.7$ is the contribution by the SFG population. Synchrotron emission is optically thin when it is produced by the shocks associated with supernovae in SFGs \citep{condon92,seymour08}.  The flattening or upturn in the number counts of radio sources seen around $S_{20cm} \leq 100-200\, \mu$Jy \citep{owen08, condon12} is explained by the emergence of this population of SFGs at faint flux density levels, exceeding those of the radio-loud AGN population that is dominant at flux densities $\geq 1\,$mJy. The increase of fractional polarization and the change of slope in the polarized number count at polarized flux densities $\leq$1~mJy also imply the increasing contribution of SFGs \citep{rudnick14}. The broad radio spectral index distributions for the GS and GN fields shown in Figure~\ref{alpha_flux} suggests the existence of both steep spectrum ($\alpha=0.5-1.0$) and flatter or inverted spectrum ($\alpha < 0.5$) sources at $S_{5GHz} < 150\, \mu$Jy, supporting the conclusions of the more recent analyses indicating that the faint $\mu$Jy radio population consist of both SFGs and radio-quiet AGN \citep{padovani09,bonzini13,rudnick14}.  A detailed study of a small sample of 14 local SFGs by \citet{klein18} has shown that there is also some scatter in the observed radio spectral index in the GHz range due to a varying degree of free-free emission and opacity effects.  What our study further indicates is that a larger sample with higher quality radio spectral index measurements are needed to characterize the relative contribution by these two populations.

\bigskip

\section{Star Formation Properties of Radio Sources \label{RADIO_SFMS}}

In the previous section, we have shown and discussed the distributions of radio spectral indices derived between 1.4 and 5~GHz from the beam matched images. In this section, we investigate how the radio spectral index correlates with star formation properties by utilizing the SFRs and stellar masses derived by the 3D-HST project \citep{skelton14, momcheva16}.

\begin{figure*}
\includegraphics[angle=0,scale=.75]{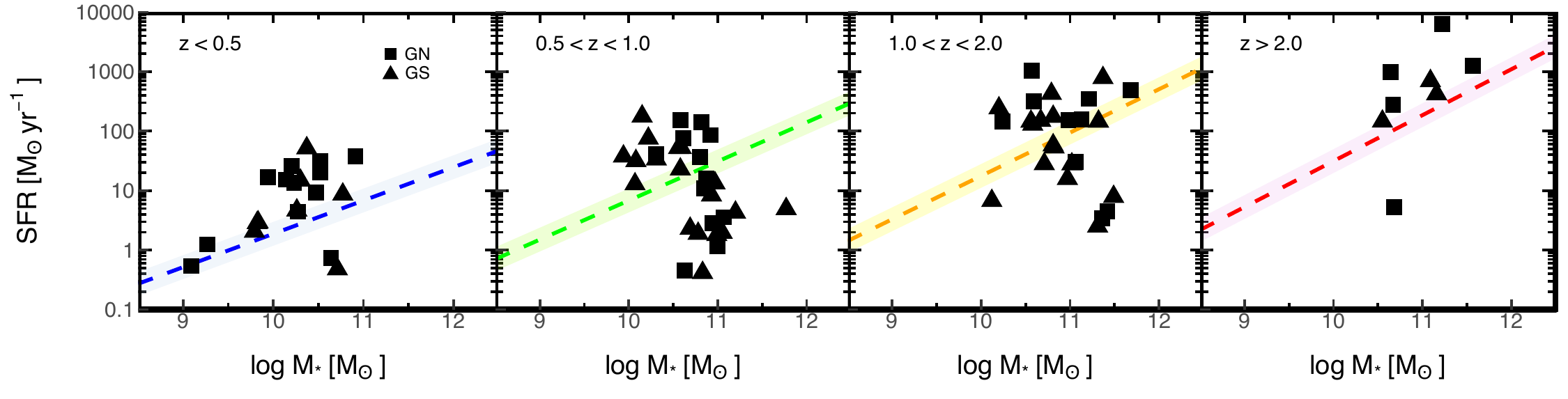}
\caption{The SFR and stellar mass distributions of the GS and GN radio sources with respect to the MS in four redshift bins. We represent the SFR and stellar mass distribution of radio sources in GN (squares) and GS (triangles) in four redshift bins, $z<0.5$, $0.5<z<1.0$, $1.0 <z<2.0$, and $z>2.0$. A dashed lines show the SFR-stellar mass relation of the MS at mean redshifts 0.25 (blue), 0.75 (green), 1.5 (orange), and 2.5 (red) with dispersion of $\pm 0.2$ (shaded regions). \label{sfr_mass_z}}
\end{figure*}

\subsection{$\Delta_{SFR}$ as a measure of SF activity \label{DELTASFR}}

The distributions of SFR and stellar mass of radio sources in GN (squares) and GS (triangles) are shown in four redshift bins in Figure~\ref{sfr_mass_z}. The dashed lines indicate the SFR-stellar mass relation of the star forming MS at a mean redshift in each panel, and the shaded regions represent dispersions of SFR-stellar mass relation at the MS with $log_{10} SFR - log_{10} SFR(MS) = \pm 0.2$ \citep{speagle14}.  As \citet{speagle14} and others noted, the MS evolves strongly with redshift, and it is not clear whether the SFRs measured at different redshifts can be compared directly in a meaningful way. A more insightful measure might be  the level of SF activity normalized by that of the MS at the same redshift.  Therefore, we define ``$\Delta_{SFR}$", the logarithm of the ratio of SFR with respect to that of the MS, as

\begin{equation}
  \Delta_{SFR} \equiv log_{10} SFR - log_{10} SFR(MS), 
\end{equation}
where $SFR(MS)$ is the $SFR$ for the star forming MS galaxy at a given stellar mass and redshift calculated using Equation (28) by \citet{speagle14}. 

\begin{figure}[h]
\center
\includegraphics[angle=0,scale=.7]{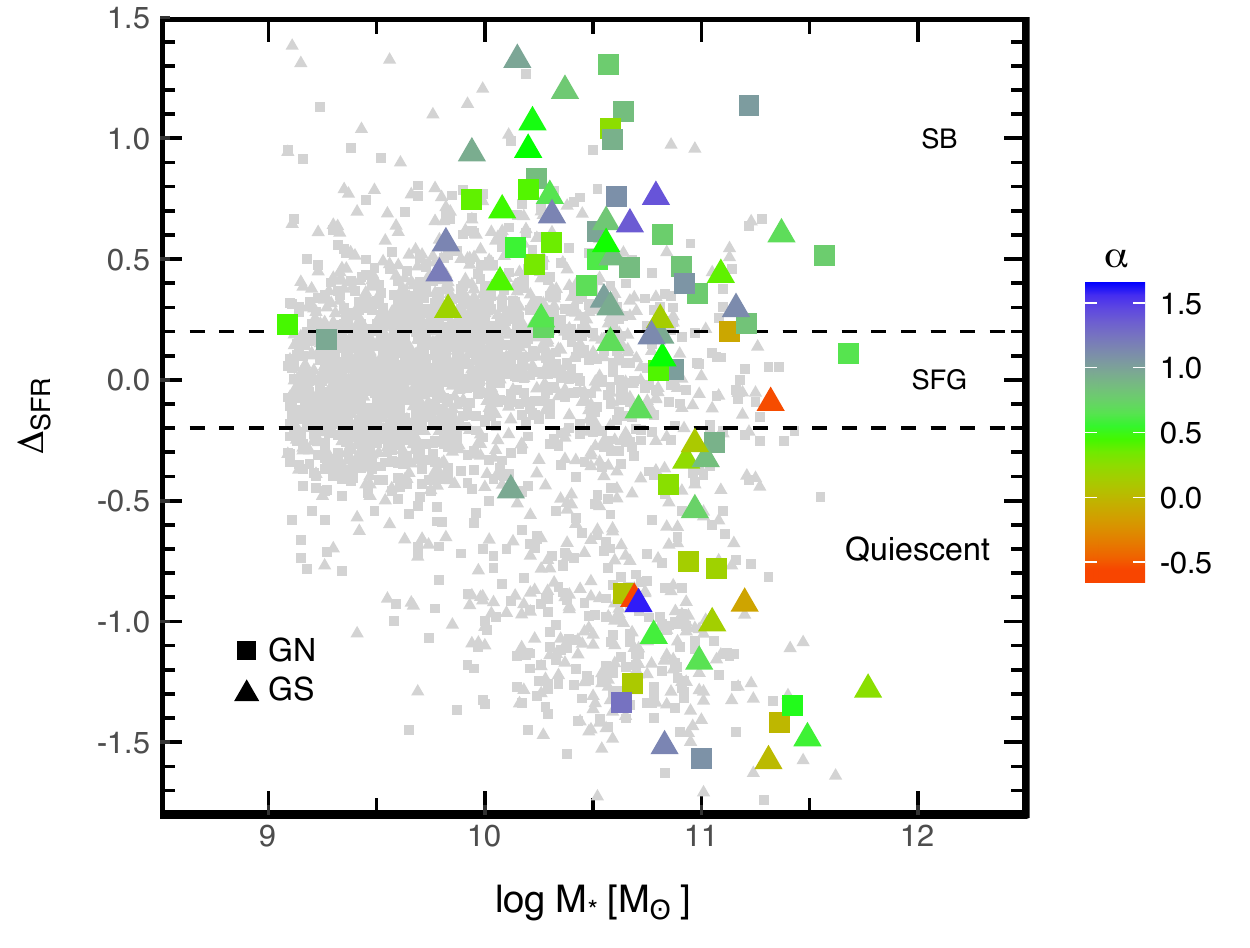}
\caption{Offset of SFR from the MS ($\Delta_{SFR}$) and stellar mass with a color code according to radio spectral index in GN (squares) and GS (triangles) fields. Small gray points are the 3D-HST galaxies without radio counterparts in GN and GS fields for a comparison. The dashed lines of $\Delta_{SFR}= \pm 0.2$ indicate the selection of SFGs. SBs are sources above a line of $\Delta_{SFR} > 0.2$ while quiescent galaxies are those below a line of $\Delta_{SFR} < -0.2$. The distribution of radio spectral index show that SFG+SB have mainly steep spectra while quiescent galaxies have flatter spectra even though both have wide distributions.  \label{sfms_selection}}
\end{figure}

Following \citet{speagle14}, we define ``SFGs" as galaxies with $-0.2 \le \Delta_{SFR} \le 0.2$, ``starbursts (SBs)" as those with $\Delta_{SFR} > 0.2$, and ``quiescent galaxies" as those with $\Delta_{SFR} < -0.2$. In total, we have 49 SBs (58\%), 10 SFGs (12\%), and 25 quiescent galaxies (30\%).  The dominance of the SB population among the $\mu$Jy radio population identified by one of the deepest surveys thus far is somewhat surprising, but this reflects the selection bias driven by the survey depth as discussed further below (also see \S~\ref{RADIO_OBS}).

In Figure~\ref{sfms_selection}, we show the distribution of $\Delta_{SFR}$ as a function of stellar mass, color-coded by radio spectral index, $\alpha$. Quiescent galaxies detected in radio continuum are on average more massive than the SFG+SB while the SFG+SB show a wider range of stellar masses as shown in Figure~\ref{sfms_selection}. The median stellar masses are 3.8$\times$10$^{10} M_{\bigodot}$ for SFG+SB and 9.3$\times$10$^{10} M_{\bigodot}$ for quiescent galaxies, respectively. The two-sided Kolmogorov-Smirnov test for two samples in R \citep{rcite} indicates that stellar mass distributions in both populations are substantially different with a p-value of $< 4.3 \times 10^{-5}$. This significant difference in mass distributions is consistent with the mass quenching scenario for quiescent galaxies \citep[e.g.,][]{kauffmann03}.

The majority of our radio sources (58\%) show intense star formation activity with $\Delta_{SFR} > 0.2$ while only 12\% of radio sources fall within the range of MS SFGs with $-0.2 < \Delta_{SFR} < 0.2$. For comparison, we show the 3D-HST galaxies without radio counterparts (light gray) in Figure~\ref{sfms_selection}. In the same stellar mass range as the radio sources (log $M_{*} \ge 9.08$), the 3D-HST galaxies undetected in radio are classified into SBs (25\%), SFGs (44\%), and quiescent galaxies (31\%). The fraction of quiescent galaxies among source undetected in radio is the same as radio detected sources.  Therefore, the main difference is in the fraction of SBs.  In all cases, the radio detected galaxies trace the high stellar mass envelope for all types of galaxies, independent of $\Delta_{SFR}$, and this is a natural consequence of a flux-limited survey as demonstrated by our selection function shown in Figure~\ref{lumz}.
Since our radio observations trace the synchrotron emission from star formation and AGN activities, these statistics imply that our radio survey is not deep enough to detect the star formation activity in the star forming MS galaxies in the full range of redshift probed, even with $\mu$Jy sensitivity.  We discuss this finding in more detail in \S~\ref{RADIO_OBS}. 

\subsection{Star Formation Activity and Radio Spectral Index \label{SFRSI}}

An apparent correlation between radio spectral index and star formation property ($\Delta_{SFR}$) is hinted in the color-coded data for radio spectral index in Figure~\ref{sfms_selection}.  Steep spectrum sources with $\alpha > 0.5$ (green and blue) appear predominantly in the $\Delta_{SFR} > -0.2$ region while sources with a flat or inverted spectrum ($\alpha < 0.5$, yellow and orange) appear mostly in the region below $\Delta_{SFR} = -0.2$.  This might be an indication that steep spectrum sources are abundant among SFG+SB galaxies with $\Delta_{SFR} > -0.2$ while few steep spectrum sources are in the quiescent galaxy region with $\Delta_{SFR} < -0.2$. 

\begin{figure}
\center
\includegraphics[angle=0,scale=.43]{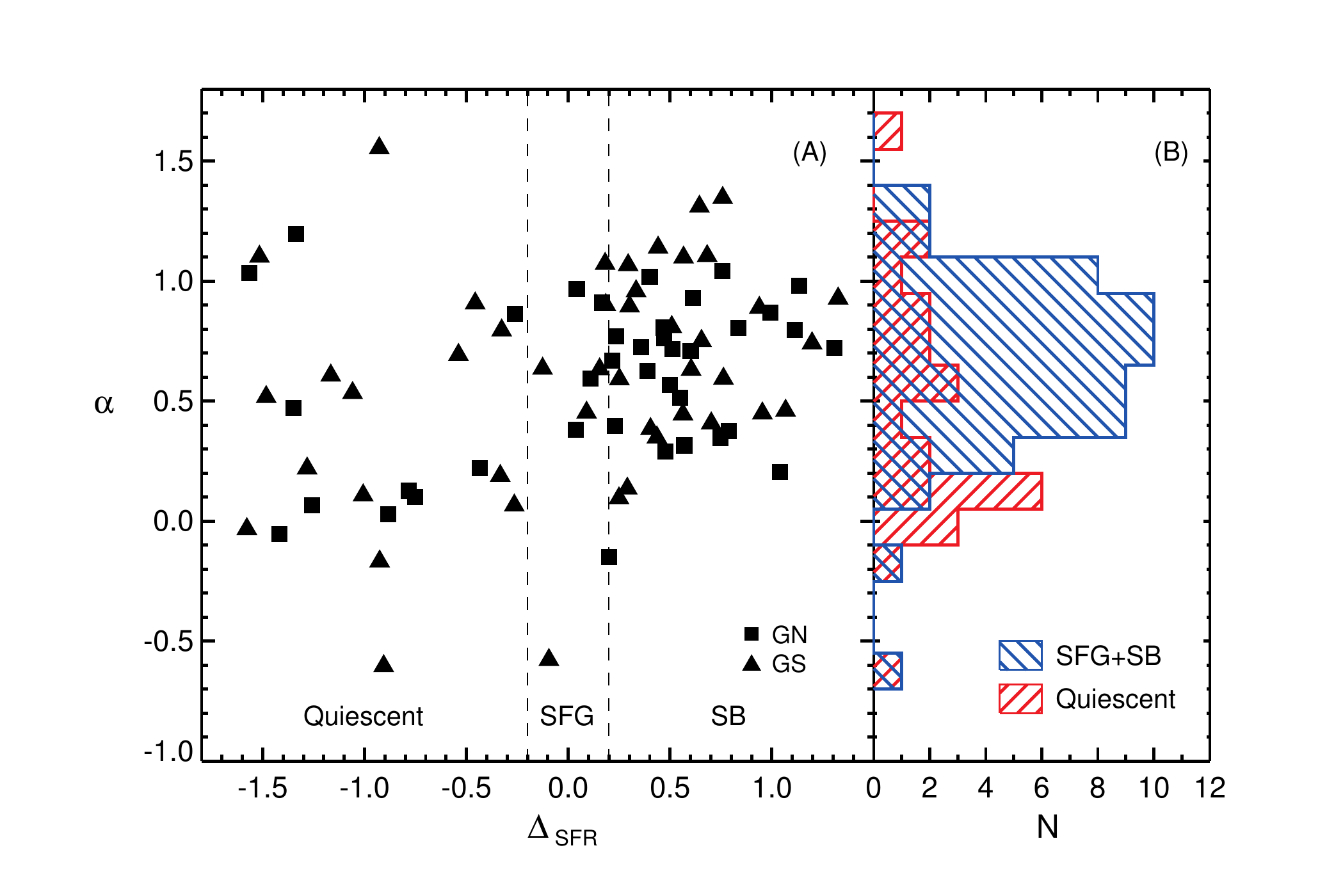}
\caption{Radio spectral index distribution as a function of $\Delta_{SFR}$. The panel (A) shows that the radio spectral index distribution of SFG+SB ($\Delta_{SFR} > -0.2$) is more tightly clustered around $\alpha \sim 0.8$, in comparison with the quiescent galaxies ($\Delta_{SFR} < -0.2$), which are distributed more uniformly and widely in spectral index $\alpha$. These trends are easily seen in the histograms of SFG+SB (blue) and quiescent galaxies (red) in panel (B). The Kolmogorov-Smirnov test indicates that the radio spectral index distributions of the two populations are different from each other with a p-value of 0.0015.  \label{alpha_delsfr}}
\end{figure}

This apparent trend is examined more directly in Figure~\ref{alpha_delsfr} by plotting the radio spectral index as a function of $\Delta_{SFR}$.  What is apparent now is that the SFG+SB galaxies are more tightly clustered around $\alpha \sim 0.8$, while the quiescent galaxies ($\Delta_{SFR} < -0.2$) are distributed more uniformly, spanning a nearly twice as large range in spectral index $\alpha$ 
-- the SFG+SB galaxies have a tighter distribution with a higher mean (0.72$\pm$0.05) than the quiescent galaxies (0.22$\pm$0.11). 
The histograms in the panel (B) of Figure~\ref{alpha_delsfr} show these trends clearly with different peak positions -- the SFG+SB galaxies (blue) have a peak at $\alpha \approx 0.8$, but the quiescent galaxies (red) have a peak at $\alpha \approx 0.13$. 
The two-sided Kolmogorov-Smirnov test for the two samples in R indicates that the null hypothesis of their radio spectral index distributions drawn from the same parent population is rejected with a p-value of 0.0015. This result is consistent with the expectation that star formation yields steep radio spectra with $\alpha \sim 0.8$ through optically thin synchrotron emission produced by supernova shocks \citep{condon92} while AGN are associated with flat or inverted radio spectra with $\alpha \ll 0.8$ through synchrotron  self-absorption \citep[e.g.,][]{debruyn76}. 

It is tempting to speculate that there is a weak trend of decreasing $\alpha$ with decreasing $\Delta_{SFR}$ if the handful of sources with $\alpha \ge1$ in the upper left corner of Figure~\ref{alpha_delsfr} are ignored.  These ultra-steep spectrum sources are generally jet-dominated AGNs, and one could separate them out morphologically, but that kind of handpicking is not generally possible for a study without the necessary spatial information.\footnote{The identification of AGN among the faint radio source population and their impact on observed properties are presented exclusively in Paper~II.}  The large spread in  $\alpha$ at a given value of  $\Delta_{SFR}$ also makes such a generalization difficult to trust.  What seems to be more certain is that this spread is real and essentially independent of star forming activity $\Delta_{SFR}$, and this has an important consequence for understanding and modeling the nature of faint radio population and their evolution, as we discuss further below.

\bigskip

\section{Radio-FIR Correlation of Radio Sources \label{QFIR_RADIO}}

The radio-FIR correlation is one of the robust indicators of star formation and black hole activities \citep{helou85,condon92,yun01,bell03}. In particular, the radio-FIR correlation of SFGs is a tight correlation with a less than 0.3 dex scatter over five orders of magnitudes in luminosity \citep{yun01}, and this obviously indicates that a strong coupling exists between dust-reprocessed emission of ultraviolet radiation from massive young stars and synchrotron radiation by cosmic rays accelerated in type II supernovae \citep{condon92}. 
In this section, we examine the radio-FIR correlation of the $\mu$Jy radio sources identified in the GN \& GS fields as a function of their star formation properties and their measured radio spectral index.

\begin{figure*}
\center
\includegraphics[angle=0,scale=.8]{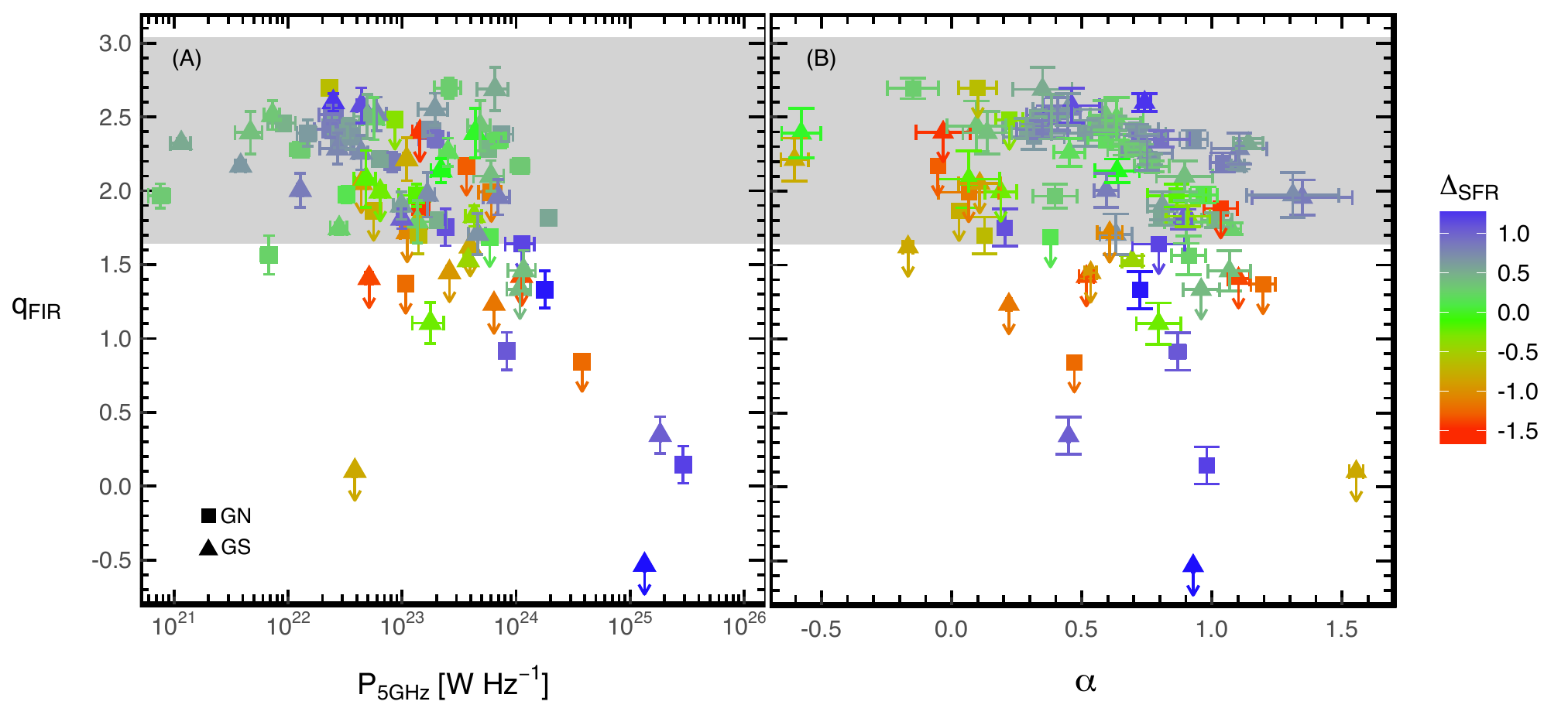}
\caption{The radio-FIR correlation parameter ($q_{FIR}$) as a function of 5~GHz power (left panel) and radio spectral index (right panel). Radio sources in GN (squares) and GS (triangles) are plotted with the color code of $\Delta_{SFR}$ and arrows showing the upper limits. A shaded region represents the range of $q_{FIR}$ of local SFGs, i.e. $q_{FIR}=$[1.64, 3.04] \citep{yun01}. Most SFG+SB (with $\Delta_{SFR} > -0.2$) are located inside the region of local SFGs while most quiescent galaxies (with $\Delta_{SFR} < -0.2$) have lower $q_{FIR}$. \label{rfc_p5}}
\end{figure*}

The rest-frame radio-FIR correlation parameter, $q_{FIR}$ is defined as 
\begin{equation}
q_{FIR} = log_{10}\left( {{L_{FIR} [W] } \over {3.75 \times 10^{12} Hz}} \right) - log_{10} P_{1.4GHz} [W Hz^{-1} ]  , 
\end{equation}
where $L_{FIR}$ is a rest-frame FIR luminosity from 40 to 120 $\mu$m \citep{helou85,yun01}. The radio-FIR correlations of radio sources as a function of redshift are shown in Figure~\ref{rfc_p5} for GN (squares) and GS (triangles), color-coded by $\Delta_{SFR}$. The overwhelming majority of the SFG+SB population (86\%) follow the local radio-FIR correlation for SFGs \citep{yun01}, and galaxies near the star-forming MS ($-0.5\le \Delta_{SFR}\le +0.5$) nearly exclusively fall within the grey band shown in the left panel of Figure~\ref{rfc_p5}.  On the other hand, only $\sim$30\% of the quiescent galaxies have $q_{FIR}$ of local SFGs, and their radio continuum emission likely has an origin other than star formation.  Most of the quiescent galaxies (76\%=19/25) are not detected in the far-IR, and they are marked with a down arrow in Figure~\ref{rfc_p5}.

A statistical analysis of the radio-FIR correlation for each subpopulation distinguished by its star formation properties shows a clear difference between the SFG+SB galaxies and the quiescent galaxies.
We have applied the Kaplan-Meier estimator for $q_{FIR}$ of the two subpopulations with the subroutine {\bf cenfit} of the statistical package NADA\footnote{Nondetects and Data Analysis for Environmental Data.} in R \citep{rcite}. This analysis shows that the SFG+SB galaxies have a median $q_{FIR}$ value of 2.26$\pm$0.09, in good agreement with the local canonical value $<q_{FIR}>\approx 2.3$ \citep{yun01}, while the quiescent galaxies have a median value of 1.10$\pm$0.10. The difference in these median values is quite substantial with a significance of $\sim 8.8 \sigma_{c}$ (the combined uncertainty for both populations is $\sigma_{c}=0.13$). 
To quantify the difference of radio-FIR correlation distributions between SFG+SB and quiescent galaxies further, we perform the Log-rank test with left-censored data using the {\bf cendiff} function in the NADA in R \citep{rcite}. This test indicates that the SFG+SB galaxies and the quiescent galaxies have entirely different distributions of $q_{FIR}$ with a p-value of $< 2 \times 10^{-6}$. These statistical tests confirm the results of previous studies that the radio-FIR correlation is a powerful tracer of star formation activity \citep{yun01,bell03}. 

An obvious trend seen in the left panel of Figure~\ref{rfc_p5} is the decreasing $q_{FIR}$ with  increasing 5 GHz radio power.  A straightforward interpretation is that radio AGN contribution is increasing both fractionally and in absolute value for the most radio luminous objects at $P_{5GHz}\ge 10^{24}$ W Hz$^{-1}$.  A somewhat surprising fact is that the majority of these ``radio-excess" objects with $P_{5GHz}\ge 10^{24}$ W Hz$^{-1}$ are also intensely starbursting galaxies with $\Delta_{SFR}\gtrsim 1$.  Similar objects found in the local Universe are mostly Seyfert AGNs associated with a nuclear starburst, but they are exceedingly rare, accounting for only 1\% of the $IRAS$ 2 Jy Sample studied by \citet{yun01}.   One might conclude a sharp increase (up to $\sim$5\%) of such AGN+SB hybrid objects at $z>1$, but our sample size is too small to be highly quantitative.  Furthermore, survey depth and sample definition might have a strong influence in such an inference as even our $\mu$Jy sensitivity is not sufficient to probe the MS star forming galaxies (see below \S~\ref{RADIO_OBS}).  Indeed, both the AGN fraction and the radio-excess fraction reported by the deeper survey of the COSMOS field by \citet{smolcic17b} are much higher, $\sim$20\%, at the $S_{1.4GHz}=50\, \mu$Jy and rising up to $\sim$50\% at $S_{1.4GHz}=100\, \mu$Jy (see their Figure~12). A similar result was also reported by a study with a different AGN identification using the VLBA observations on the same field, where the AGN fraction is $>$40$-$55\% at 100 $< S_{1.4GHz} < $ 500 $\mu$Jy \citep{herrera-ruiz18}. 

The dependence of radio-FIR correlation on radio spectral index is examined on the right panel of Figure~\ref{rfc_p5}, and the quiescent galaxies with $\Delta_{SFR}\le$ -0.2 show systematically lower $q_{FIR}$ (on average by 0.6-0.8) compared with the SFG+SB population, nearly independent of radio spectral index $\alpha$.  An in-depth analysis of the similarities and differences among these different subpopulations is discussed in our next paper (Paper~II), but this is another indication that quiescent galaxies are indeed a distinct population in their radio and IR properties as well.  It is interesting that the extreme steep spectrum quiescent galaxies identified in Figure~\ref{alpha_delsfr} and discussed in \S~\ref{SFRSI} are {\em not} extreme outliers and instead nearly follow the normal radio-FIR correlation.  A real outlier in the distribution is again the radio-excess SBs with $\Delta_{SFR}\gtrsim 1$ discussed above, and their radio spectral index is typically around $\alpha\sim +0.9$, indistinguishable from the bulk of the normal SFGs and SBs.  Intense starbursts associated with massive galaxies in the local universe, such as luminous infrared galaxies (LIRGs) and ultraluminous infrared galaxies (ULIRGs), are associated with high free-free opacity, leading to the flattening of radio spectrum \citep[e.g.,][]{klein18} and even obscuring a radio AGN altogether at longer wavelengths (e.g., Mrk~231).  Therefore, the distribution and geometry of starburst activity in these $z>1$ luminous radio-excess SBs are somehow different from local examples.  And they certainly cannot be identified from their luminosity and radio spectral index alone.  Future higher resolution observations that can resolve the star-forming structures and kinematics are required to yield deeper insight on these sources.

\bigskip

\section{Discussion \label{DISCUSSION}}

\subsection{Importance of Survey Sensitivity \label{RADIO_OBS}}

What makes deep radio continuum imaging attractive as a tool for studying galaxy evolution is the high angular resolution of an interferometer like the VLA to deliver spatial information at much better than 1\arcsec, free from the fundamental limits of source confusion that restrict the usefulness of current infrared facilities such as {\em Herschel}.  Advances in sensitivity through increased bandwidth and collecting area also enable us to probe star forming galaxies and AGN population at cosmological distances directly.  One of the main goals of this VLA study of the GOODS cosmology fields is to analyze the nature of the faintest radio continuum sources detectable with the current technology and establish technical specifications for future surveys for galaxy evolution using facilities such as MeerKAT, ASKAP, and eventually the Square Killimeter Array.

The plot of rest-frame 5 GHz radio power versus spectroscopic redshift shown in Figure~\ref{lumz} and the analysis of their star formation properties discussed in \S~\ref{RADIO_SFMS} clearly demonstrate that our deep 5 GHz continuum data indeed probes star forming galaxies out to $z\sim 3$.  On the other hand, our detailed examination of their specific star formation rate shown in Figure~\ref{sfms_selection} finds that the fraction of SBs (58\%) in our radio sources are more than twice the fraction among the parent general galaxy population in the 3D-HST survey.  Since there are no reasons for radio-selected SFGs to be fundamentally different from optical or UV selected SFGs, this statistical difference is likely the result of the combined effects of our survey depth and the strong evolution of cosmic star formation rate density \citep[see review by][]{madau14}.
  
\begin{figure}[t]
\center
\includegraphics[angle=0,scale=.71]{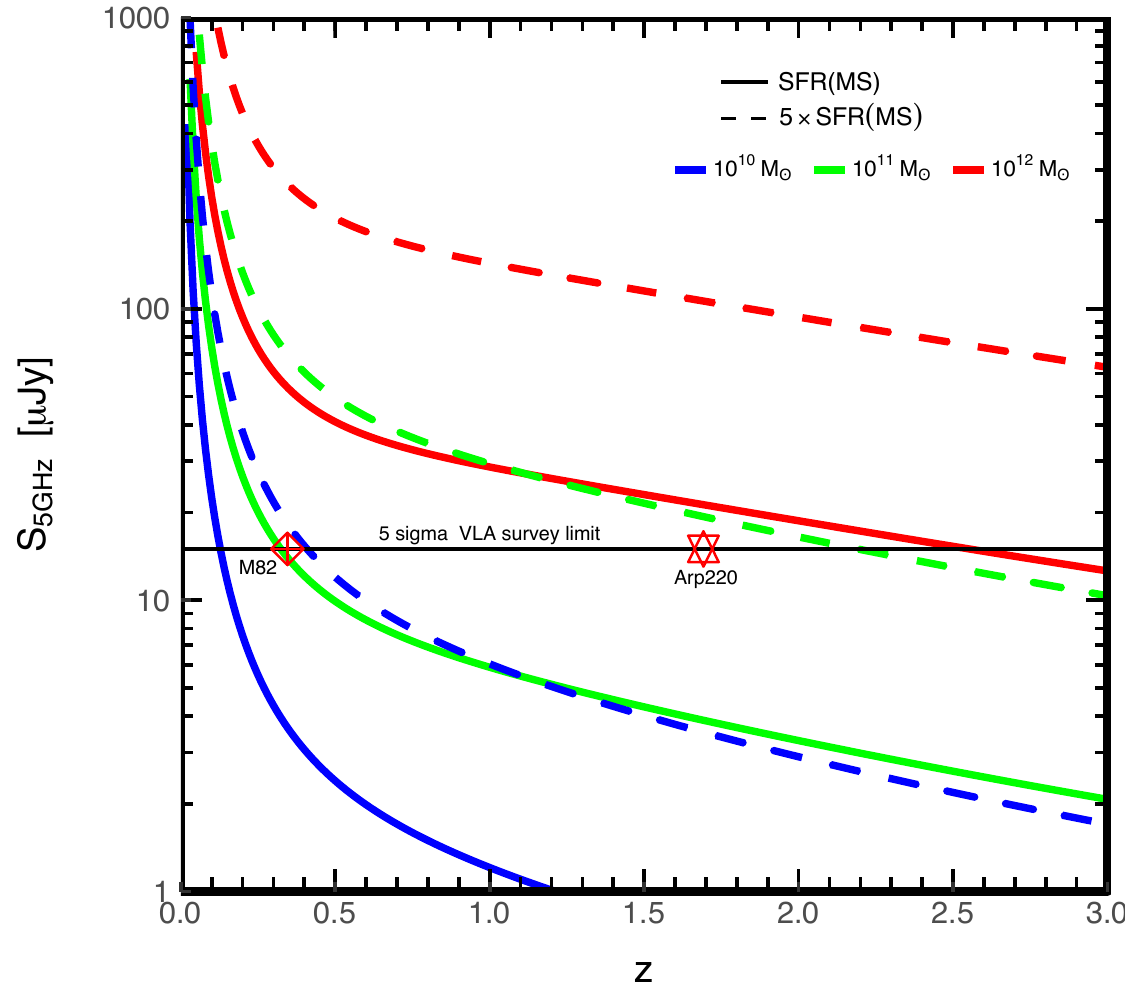}
\caption{Detectability of MS SFGs and a sensitivity of radio observations. We show the observable galaxies with a certain SFR and stellar mass as a function of redshift. We show SFGs with SFR of MS (solid lines) and $5 \times$SFR of MS (dashed lines) as a function of redshift with respect to the stellar masses of $10^{10} M_{\odot}$ (blue), $10^{11} M_{\odot}$ (green), and $10^{12} M_{\odot}$ (red). The survey limits ($5\sigma$) of our radio observations are indicated by the horizontal lines, i.e. 15$\mu$Jy for GS. As examples, we marked the maximum redshifts of detecting M82-like (red diamond) and Arp220-like (red star) galaxies at the survey limits. \label{detection}}
\end{figure}

To explore this further, we show the calculated 5~GHz radio flux density of SFGs with SFR of MS (solid lines) and $5 \times$SFR (dashed lines) for stellar masses of $10^{10} M_{\odot}$ (blue), $10^{11} M_{\odot}$ (green), and $10^{12} M_{\odot}$ (red) in Figure~\ref{detection}. We assume that SFR scales with 1.4~GHz radio power following the radio-total IR correlation with $q_{TIR}=2.64$ \citep{murphy11} and a single average radio spectral index of +0.8 (but see the discussion on potential bias below). In general, angular resolution and source size are important considerations for survey sensitivity.   Here, we make a simplifying assumption that most sources detected in a deep survey like this are at high redshifts are unresolved or marginally resolved \citep{owen08, murphy17, owen18}.\footnote{Median of radio source sizes reported at 1.4~GHz by \citet{owen08} and \citet{owen18} are 1.2\arcsec - 1.5\arcsec while the median source size at 10~GHz reported by \citet{murphy17} is 0.17\arcsec~$\pm$ 0.03\arcsec. The apparent difference in these median radio source sizes is likely attributable to the structures present in these radio sources and the differences in the surface brightness sensitivity achieved.} At our 15$\, \mu$Jy ($5\sigma$) survey limit for the GS field (black horizontal line), the maximum observable redshifts for star forming MS galaxies (dashed lines) are $z$=0.13 for $10^{10} M_{\odot}$ (solid blue), $z$=0.32 for $10^{11} M_{\odot}$ (solid green), and $z$=2.55 for $10^{12} M_{\odot}$ (solid red).  SFGs with $5\times$SFR of the MS can be detected out to $z$=0.41 for $10^{10} M_{\odot}$ (dashed blue), $z$=2.19 for $10^{11} M_{\odot}$ (dashed green), and $z >$3 for $10^{12} M_{\odot}$ (dashed red). 
In terms of well-known local SFGs, we can detect M82-like galaxy out to $z$=0.34 and Arp220-like galaxy out to $z$=1.63, respectively. Therefore, even with the $\mu$Jy sensitivity we achieved in these two GOODS fields, we can probe a main sequence SFG with a stellar mass of $10^{11} M_{\odot}$ only out to $z\sim$0.3, and our survey is strongly biased to ULIRG-like starbursts and AGN-host galaxies at $z>1$. 

This same plot also demonstrates that directly probing the evolution of the star forming MS galaxies will require a {\em much} deeper survey.   To probe a MS SFG with $SFR=10 M_{\odot}$ yr$^{-1}$ at the Cosmic Noon ($z=2.5$) at $5\sigma$, a 5~GHz radio survey needs to reach a survey sensitivity of 28~nJy with the Next Generation VLA or Square Kilometer Array. This required sensitivity is $\sim$11.5 times deeper than the existing deepest 5~GHz continuum survey of the Hubble Ultra Deep Field by \citet{rujopakarn16} and {\em more than 100 times deeper} than our own surveys presented here.  

\medskip
\subsection{Importance of Angular Resolution \label{RESOLUTION}}

In the previous section, we discussed the importance of sensitivity in probing star forming galaxies at cosmological distances and the requirement for future surveys to improve the sensitivity by more than an order of  magnitude to probe the evolution of the main sequence SFGs.  However, another surprising outcomes of our deep VLA 5 GHz surveys is that simply obtaining a deeper data itself does not guarantee probing much deeper into the luminosity function.  As discussed in \S~\ref{PREVOBS}, the comparison of the past and recent deep surveys seems to suggest that the rise in source density is {\em apparently} much flatter than the Euclidean case.  Obviously this is not an entirely fair and rigorous comparison, and the situation is quite a bit more complex.

\begin{figure}[t]
\center
\includegraphics[angle=0,scale=.7]{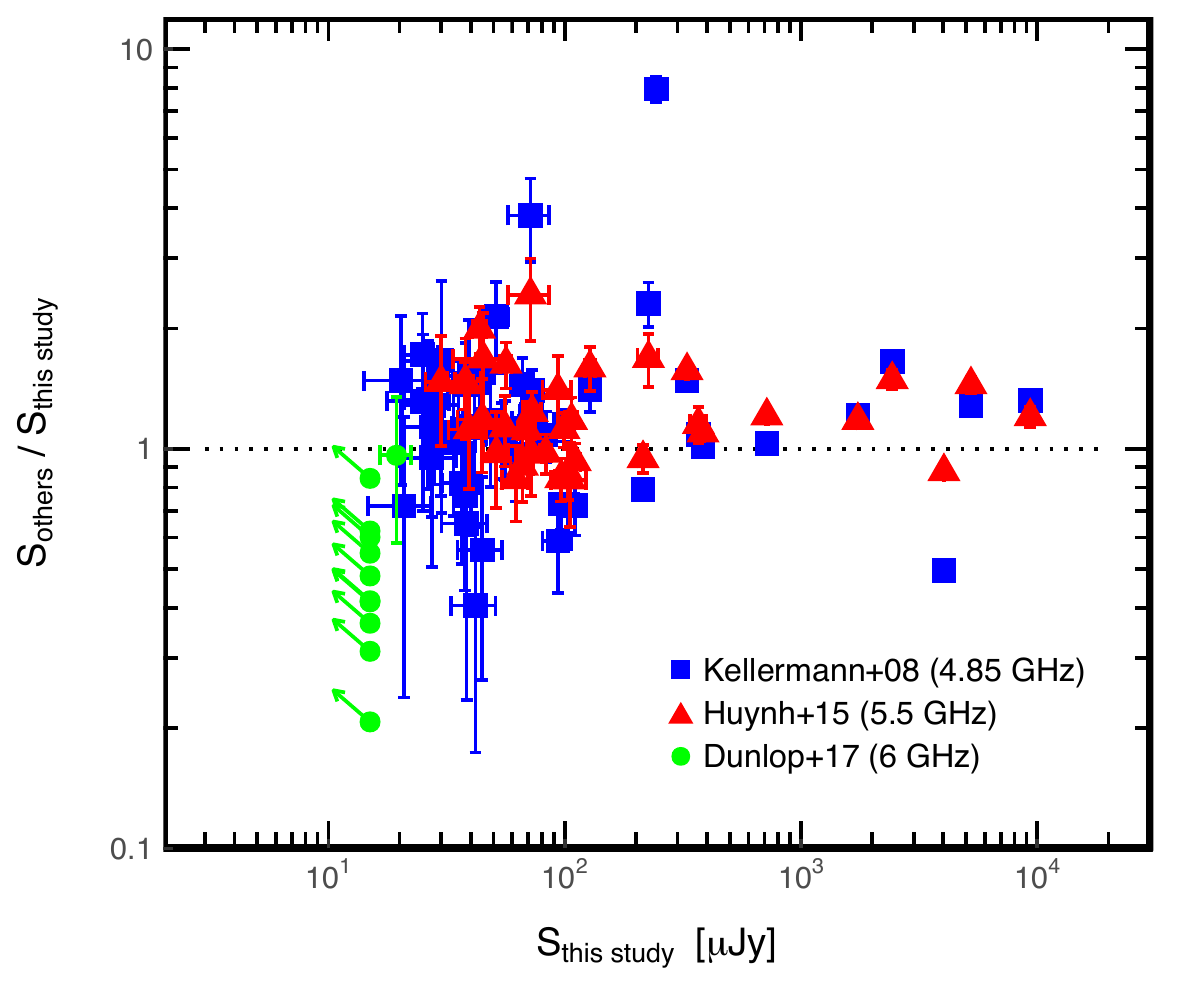}
\caption{Measured flux density comparison among the radio sources in the GS field with those reported by previous studies with different angular resolution.  Those by \citet{kellermann08} and \citet{huynh15} with $\sim3$ times larger beams are on average $\sim$30 percent larger.  The higher resolution survey by \citet{rujopakarn16} with 0.61\arcsec$\times$0.31\arcsec beam has only one detected source in common \citep[the 6 GHz source flux densities are actually reported by][]{dunlop17} that agrees well with ours.  The dotted line is the unity ratio line to guide the eye.
\label{GSCompare}}
\end{figure}

A potentially important experimental parameter here is angular resolution.  Both statistical \citep[e.g.,][]{windhorst90,morrison10} and direct imaging \citep[e.g.,][]{chapman04} studies have shown that faint radio sources have an intrinsic size of  $1\arcsec-2\arcsec$.  Resolving sources with an angular resolution higher than the intrinsic size can negatively impact deep surveys of star forming galaxies in two ways: (1) by fragmenting individual radio sources into multiple components, especially in the low SNR regime; and (2) loss of surface brightness sensitivity and the resulting loss of extended emission.  The former is a well-known phenomenon for nearly all deep radio surveys, and most previous studies have produced catalogs of ``source components" as well as integrated source catalogs.   In analyzing the $0.5\arcsec$ imaging data of the GN field, \citet{guidetti17} identified the loss of surface brightness sensitivity and their bias toward compact sources as the primary cause for their extra-ordinarily high AGN fraction.  Only a modest (a factor of $\sim3$) increase in the source density reported by \citet{rujopakarn16} in their ultra-deep imaging of the GS field with nearly 10 times better sensitivity than our survey is likely driven by the loss of flux density and surface brightness sensitivity resulting from their using very high angular resolution (0.61\arcsec$\times$0.31\arcsec).   

We explore the impact of angular resolution on flux recovery further by comparing the measured flux density of the faint radio sources in the GS field reported by different surveys with varying angular resolution in Figure~\ref{GSCompare}.  The flux densities reported by \citet{kellermann08} at 4.85~GHz and by \citet{huynh15} at 5.5~GHz were both measured using a $\theta\approx4\arcsec$ beam, and these flux densities are systematically higher when compared with our measurements obtained with a 1.5\arcsec\ beam. The average flux ratio between the \citet{kellermann08} flux density to our flux density is 1.34, with a median ratio of 1.14.  Similarly, the average and median ratios of the \citet{huynh15} flux density to our flux density is 1.26 and 1.18, respectively.  A small correction due to intrinsic spectral index is neglected, as both low angular resolution measurements are significantly larger (about 30\%) than our measurements with an effective center frequency of 5.25 GHz.  These measured differences are much larger than the expected absolute calibration uncertainties ($\lesssim$10\%) associated with the standard flux density bootstrapping calibration.  The comparison with the higher resolution (0.61\arcsec$\times$0.31\arcsec) imaging by \citet{rujopakarn16} does not provide much new insight as there is only one source in common.    

In summary, observing angular resolution smaller than the expected intrisic radio source size of $1\arcsec-2\arcsec$ can lead to a significant systematic bias in deep radio surveys.
Carefully accounting for this resolution effect and surface brightness sensitivity is an important consideration for all future ultra-deep surveys with nJy sensitivity.

\medskip
\subsection{Importance of Accurate Radio Spectral Index  \label{DISC_SI}}

Obtaining accurate radio spectral indices is an important step in studying the radio-FIR correlation and its evolution over the cosmic time because computing the rest-frame radio-FIR correlation requires a correction with a ``$log_{10} \left[(1+z)^{1-\alpha} \right]$" dependence on radio spectral index, associated with the $k$-correction for the observed radio power.  This has the largest impact at the highest redshifts, where the evidence for any evolution in the radio-FIR correlation is expected to be the most pronounced.  

Many previous studies of faint radio source population have applied only a partial correction for this spectral index effect, largely because of practical constraints, but the magnitude of the resulting error may have been under-appreciated. Ideally, one should obtain observations at two different frequencies with matched beams and depths to derive correct radio spectral index.  However, conducting observations in {\em two} frequency bands can be prohibitively expensive in telescope time, especially for deep surveys that require tens to hundreds of hours of integration time in each band.  Instead, a common practice is to take advantage of existing survey at another frequency, as we have done using the existing 1.4 GHz surveys by \citet{miller13} and by \citet{owen18}.  If the complementary archival data are not readily available in raw format as is often the case, however, radio spectral index has to be computed without the beam correction \citep[e.g.,][]{ivison10a, bourne11, magnelli15, delhaize17}.  Alternatively, a number of other studies have resorted to adopting a single average radio spectral index of 0.7-0.8 instead \citep[e.g.,][]{appleton04, ibar08, murphy09a, sargent10, ivison10b, mao11}. Because even SFGs at $z\ge1$ are resolved at $\sim1\arcsec$ scales, ignoring this resolution effect can lead to significant systematic errors in computing the total radio power and the radio spectral index.  Similarly, the radio spectral index distribution is intrinsically broad as discussed in \S~\ref{ALPHA}, and adopting a single value of $\alpha$ can introduce significant errors in the derived source properties.  Here, we analyze both of these issues quantitatively using our GN and GS deep survey data with and without the appropriate corrections.

\begin{figure*}
\center
\includegraphics[angle=0,scale=0.85]{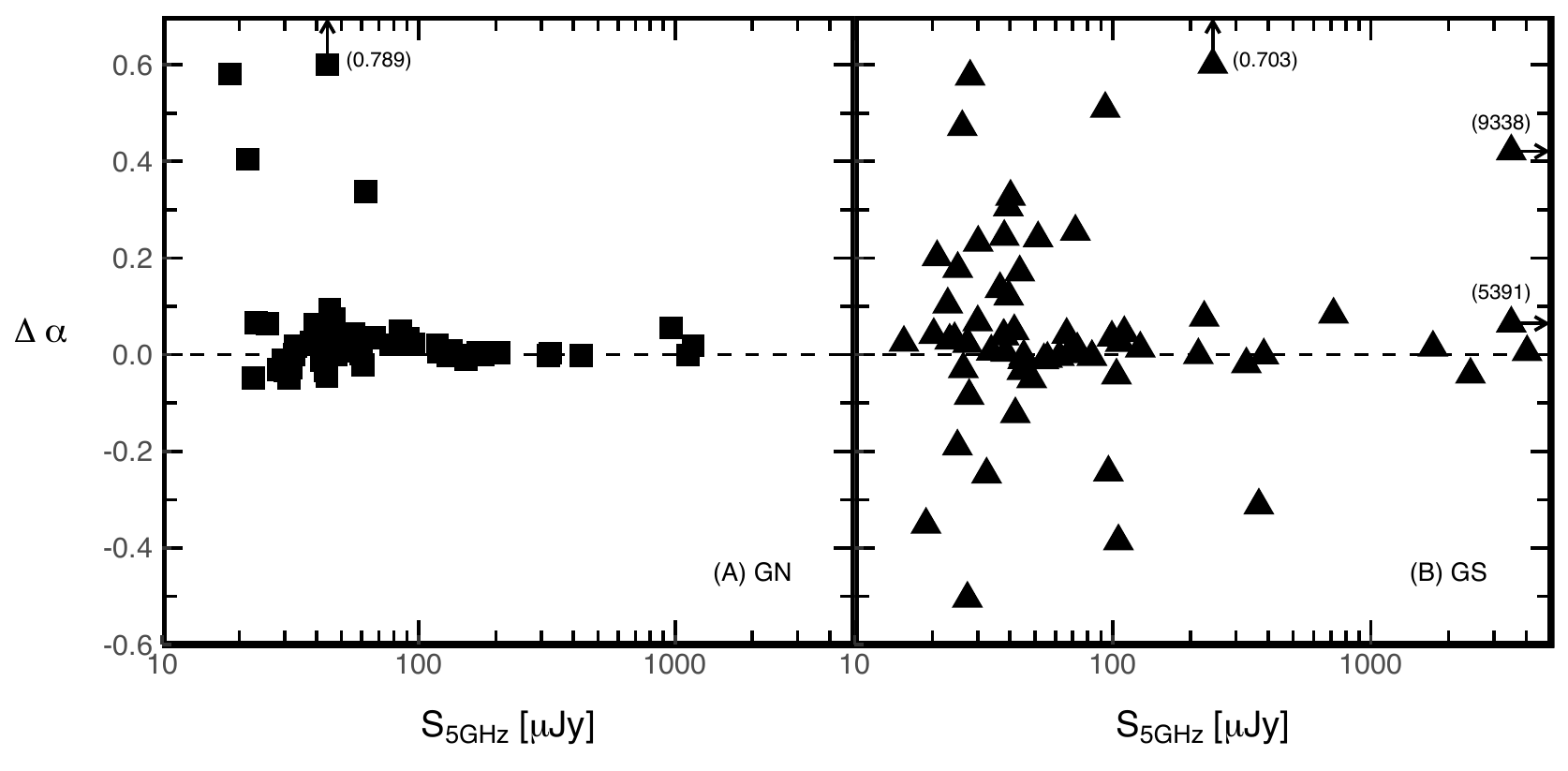}
\caption{Deviations of radio spectral index measured without matching the beam sizes ($\alpha_{non}$) from that measured by matching beam sizes ($\alpha_{beam}$).  The deviation of radio spectral index ($\Delta \alpha=\alpha_{non} - \alpha_{beam}$) is shown as a function of 5~GHz flux density for the GN field (left panel) and for the GS field (right panel).  The main difference is the much larger synthesized beam for the 1.4 GHz data in the GS field.
\label{alpha_diff_flux}}
\end{figure*}

\subsubsection{Importance of Beam-matching for the Radio Spectral Index Calculation \label{NON_ALPHA}}

A measured radio spectral index is a direct indicator of the primary radiation mechanism for the observed radio power.   In this section, we compare the radio spectral index estimated without matching beam sizes ($\alpha_{non}$) and with those with matched beams ($\alpha_{beam}$), to quantify the importance of the beam effect. The ratio of beam areas is mostly between 1.2 and 1.9 for the GN sources while the GS sources have an average beam area ratio of 10.2, requiring a much larger correction. 

The impact of ignoring the beam size difference is clearly shown in the plot of the deviation of radio spectral index $\alpha_{non}$ from $\alpha_{beam}$ ($\Delta \alpha \equiv \alpha_{non} - \alpha_{beam}$) as a function of total 5 GHz flux density in Figure~\ref{alpha_diff_flux}.  In the GN field (left panel) where the synthesized beams of 5 GHz and 1.4 GHz data are closely matched, the change is small for most objects as expected.  A few sources still show a large deviation with a large positive $\Delta \alpha$ value, indicating that extended or blended sources can lead to large errors in derived spectral indices even when the beam size difference is relatively small. 
Otherwise the observed scatter is consistent with the expected increase in the noise of the 5 GHz data by the larger photometry aperture.  The scatter in the derived spectral index is much larger in the GS field (right panel), and this reflects the impact of a much larger beam difference. As in the GN field, the source distribution is biased to the large positive $\Delta \alpha$ values with a mean of $0.054$, especially among $S_{5GHz}\ge$ 1 mJy sources that are usually associated with extended radio jet sources.  

This analysis clearly demonstrates that a small but non-negligible fraction of radio sources are resolved at 1\arcsec\ scale by our 5 GHz beam, and beam-matching is critically important in deriving a correct radio spectral index.  This analysis also indicates that our deep 5 GHz data might suffer from loss of flux density due to spatial filtering, even after the beams are matched by smoothing. These combined effects lead to a systematic bias to a steeper (more positive) spectral index and smearing of the overall spectral index distribution, as seen in Figure~\ref{alpha_flux} and discussed in section \S~\ref{ALPHA}. Indeed, all interferometric observations are subject to loss of flux density, and matching the resolution to source size is the best that can be done without obtaining additional data.

\bigskip

\begin{figure*}
\center
\includegraphics[angle=0,scale=0.75]{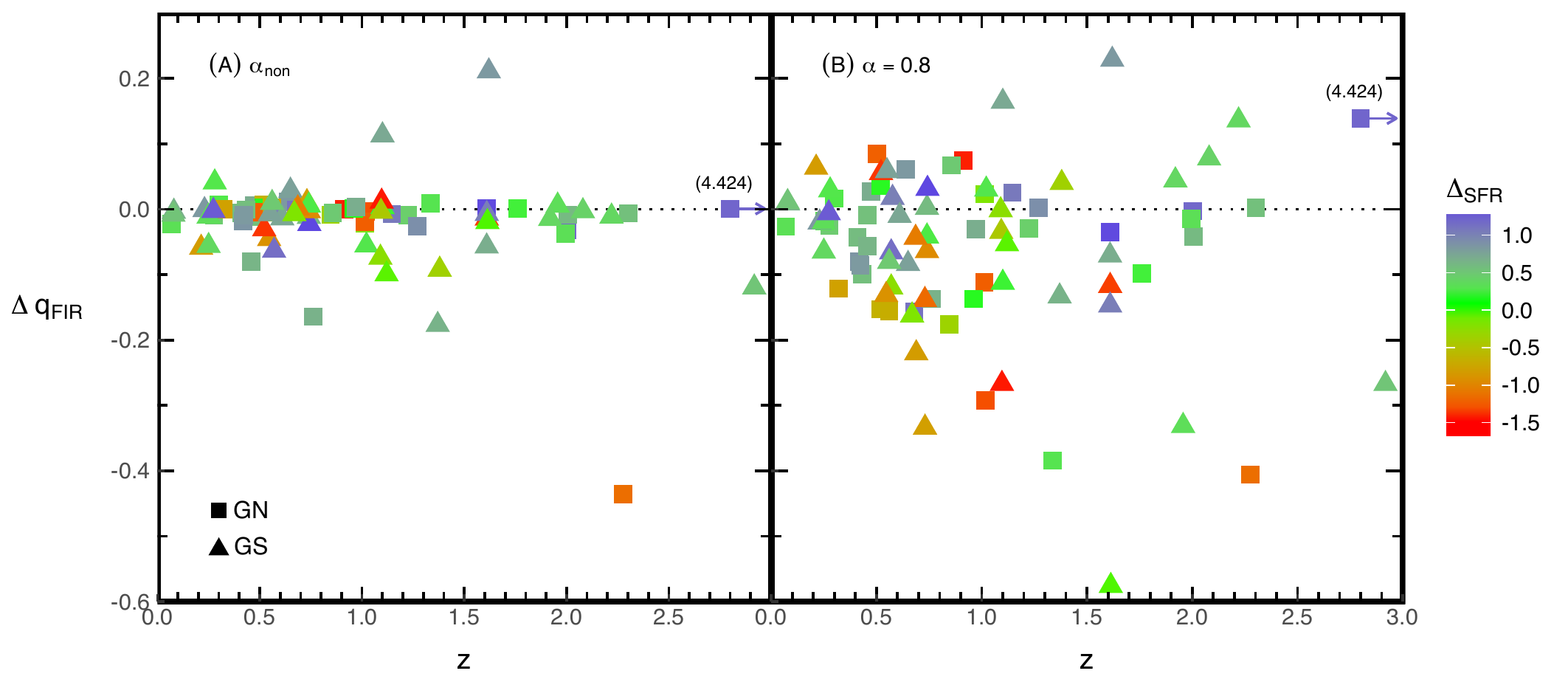}
\caption{Deviations of derived radio-FIR correlations resulting from errors of radio spectral index. The deviation of radio-FIR correlation parameter is shown as a function of redshift, where the deviations of radio-FIR correlations are originated by the radio spectral index by unmatched beam, $\Delta q_{FIR} = q_{FIR}(\alpha_{non})-q_{FIR}(\alpha_{beam})$ in left panel and by adopting a single value of $\alpha=0.8$, $\Delta q_{FIR} = q_{FIR}(\alpha=0.8)-q_{FIR}(\alpha_{beam})$ in right panel.
\label{qdiff}}
\end{figure*}

\subsubsection{Impact of Spectral Index on Radio-FIR Correlation}
\medskip

The rest-frame radio-FIR correlation depends on the radio spectral index through the k-correction for the rest-frame radio power, and there are two common ways which incorrect radio spectral index has impacted the radio-FIR correlation analysis in the literature: (a) not matching beams; and (b) adopting a single value of $\alpha$.  Here, we demonstrate how both of these errors in radio spectral index can lead to systematic deviations in the derived radio-FIR correlation parameters $q_{FIR}$ using our data. The deviation of radio-FIR correlation is defined as $\Delta q_{FIR} \equiv q_{FIR} (\alpha_{non}) - q_{FIR} (\alpha_{beam})$ [for the unmatched beam case], and they are shown as a function of redshift, color-coded by $\Delta_{SFR}$, in Figure~\ref{qdiff}.

As discussed in the previous section, the net effect of not correcting for the beam size difference is over-estimating radio spectral indices (for this study, because of the higher angular resolution of the 5 GHz data), which in turn leads to a larger k-correction and an over-estimation of the rest frame radio power.  As shown on the left panel of Figure~\ref{qdiff}, the overall scatter in $\Delta q_{FIR}$ resulting from not matching the beams is not large, less than 0.1 in dex. However, all sources with a significant deviation in $q_{FIR}$ are {\em nearly uniformly and systematically towards a lower value with a mean scatter of -0.019, and this bias is larger in magnitude at a higher redshift} because of a larger k-correction.

The common practice of adopting a single ``average" value (e.g., $\alpha=0.8$) leads to an even greater scatter and a stronger bias in $q_{FIR}$ than the unmatched beam case, as shown on the right panel of Figure~\ref{qdiff}.  The magnitude of the scatter in $\Delta q_{FIR}$ is now nearly 0.2 in dex, approaching the {\em total} intrinsic scatter in the observed radio-FIR correlation for the local SFGs  \citep{yun01}. In addition, $\Delta q_{FIR}$ is heavily biased towards the negative values with a mean of -0.061 (and growing with redshift), as is the case for the unmatched beam.  Both of these trends are the direct results of the large and asymmetric spread in the measured radio spectral index distribution shown in Figure~\ref{alpha_flux}.

The fact that both of these common errors in radio spectral index can lead to a significant scatter and a strong bias in the derived $q_{FIR}$ is a serious concern for the study of the faint radio source population in general and the study of radio-FIR correlation specifically.  The magnitude of the error grows systematically with redshift and is more biased to a lower value of $q_{FIR}$, and this has an important consequence for the evaluation of possible evolution of the radio-FIR correlation.   We will discuss this effect in the context of radio-FIR correlation evolution in Paper II.

\bigskip
\section{Conclusions \label{CONCLUSION}}

We reported the first results from our deep and wide VLA 5~GHz surveys of the GN and GS fields with the resolution and sensitivity of $\theta=1.47\arcsec\times1.42\arcsec$ \& $\sigma=3.5\, \mu$Jy beam$^{-1}$ and $\theta=0.98\arcsec \times0.45\arcsec$ \& $\sigma=3.0\, \mu$Jy beam$^{-1}$, respectively.  The central deep cosmology fields with HST and other multi-wavelength data are covered with a nearly uniform sensitivity and resolution, and a total of 52 \& 88 sources are identified at $\ge5\sigma$ significance in the 109 \& 190  arcmin$^{2}$ survey areas, respectively.  We have carefully derived their radio spectral indices by utilizing the existing 1.4~GHz images and catalogs by \citet{owen18} and by \citet{miller13} and examined the radio spectral index distribution and radio-FIR correlation using only a subset of 84 sources with a reliable spectroscopic redshift to minimize introducing additional scatter.   Some of the main results from our analyses of these data include:

\begin{enumerate}
  \item The radio spectral index is measured from beam-matched images of 1.4 \& 5~GHz, and its distributions show the clustering of faint radio sources with $S_{5GHz} \lesssim 150 \mu$Jy at around the steep radio spectral index of $\alpha \sim$0.8, which has not seen in previous studies.  The associated peak in the GN field is more distinct than in the GS field where the distribution is more smeared out by higher noise.   The overall spectral index distribution derived is quite broad, ranging $-0.5 \le \alpha \le 1.4$, as many earlier studies have reported.    
  
  \item The star formation activity is characterized by the distance from the ``star formation main sequence" \citep{speagle14}, taking into account the strong evolution of SFR with redshift.  The majority of faint radio sources are identified as SBs (58\%) while only 12\% is identified as star forming MS galaxies with $|\Delta_{SFR}| \le 0.2$.  The remaining 30\% are quiescent galaxies with $\Delta_{SFR} \le -0.2$.  This high frequency of SBs is traced to the relatively poor sensitivity of even this deep continuum survey to normal MS SFGs at $z\ge 0.5$, and {\em future surveys with up to 100 times better sensitivity ($\sigma_{5GHz} \lesssim 30$ nJy) are needed in order to trace the evolution of the star forming MS at the Cosmic Noon ($z=2.5$).}   Our comparison of flux density measurements and source density at different angular resolution support the $\sim$1\arcsec\ extent of intrinsic radio source size reported by previous studies \citep[e.g.,][]{windhorst90,chapman04,morrison10}, and future ultra-deep surveys should carefully consider the resolution effects, e.g., such as surface brightness sensitivity as well.
  
  \item The SFG+SB population shows a significantly tighter distribution of spectral index than the quiescent galaxies, as shown in Figure~\ref{alpha_delsfr}, suggesting a systematically different origin for their radio emission.  The overwhelming majority of the SFG+SB population (86\%) follow the local radio-FIR correlation for SFGs \citep{yun01} with a median $q_{FIR}$ value of $2.26\pm0.09$.  Only $\sim$30\% of quiescent galaxies follow the same trend, with a median $q_{FIR}$ value of $1.10\pm0.10$ -- most of the quiescent galaxies (76\%) are not detected in any of the $Herschel$ far-IR bands.  The fraction of radio-excess objects with $q_{FIR} \le 1.6$ increases with increasing 5~GHz radio power, especially for objects at $z\ge1$ with $P_{5GHz}\ge 10^{24}$ W Hz$^{-1}$, and the majority of these objects are intense starburst galaxies with $\Delta_{SFR}\gtrsim1$.  This may indicate a sharp rise in the AGN+SB hybrid population at these redshifts, as suggested by previous studies.

 \item Determining and applying correct radio spectral indices is important for deriving accurate radio power and analyzing the radio-FIR correlation.  Using our own survey data, we demonstrate that the common practice of not matching the beams carefully can lead to a significant and strongly bias estimation of $\alpha$ and over-estimation of radio power for high redshift sources.  More importantly, as shown in Figure~\ref{qdiff},  the widely used practice of adopting a single ``characteristic" value of spectral index ($\alpha \approx 0.7-0.8$) leads to a much greater scatter matching or exceeding the intrinsic scatter seen in the local population and also a strong systematic bias to negative $q_{FIR}$ values, resulting from the broad width and the asymmetry in the intrinsic radio spectral index distribution. 
  
\end{enumerate}

Lastly, analyzing our data using the photometric redshifts from the 3D-HST project leads to an additional scatter of 0.112 dex in the derived radio-FIR correlation -- see Appendix~\ref{ZPHOT}.  The resulting scatter is nearly symmetric, unlike the errors in spectral index discussed above, and analyzing a much larger sample with high quality photometric redshifts might be acceptable for future studies requiring much better statistics.

\acknowledgments

We are grateful to Ryan Cybulski, St\'{e}phane Arnouts, Olivier Ilbert for the use of Le Phare, Katherine E. Whitaker for help in using the 3D-HST, and Daniel Q Wang for a discussion of X-ray AGN and HMXBs. We also thank Urvashi Rau for a discussion about the radio imaging and Ken Kellermann for a valuable discussion. 
Hansung B. Gim acknowledges special thanks to NRAO employees for their hospitality when he was visiting NRAO at Socorro, NM, and for valuable helps offered through the NRAO helpdesk.
We appreciate the anonymous referees to help us improving this paper. \\

\facility{Very Large Array, Herschel Space Observatory, Hubble Space Telescope (3D-HST)} 

\software{CASA, AIPS, R, IDL, Python}

\bigskip
\bigskip

\appendix

\section{A. Spectroscopic Redshifts versus Photometric Redshifts \label{ZPHOT}}

\begin{figure*}[h]
\center
\includegraphics[angle=0,scale=.80]{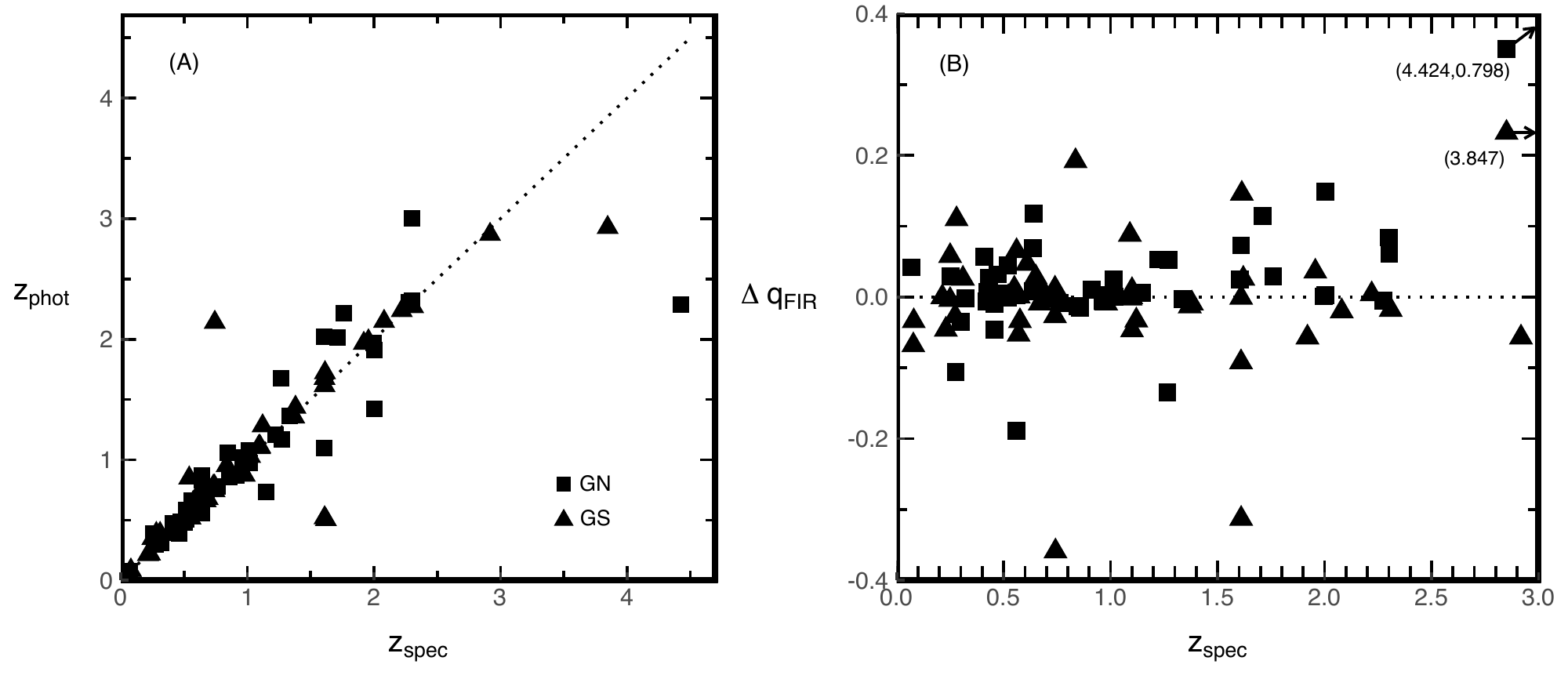}
\caption{Comparison of the photometric and spectroscopic redshifts from the 3D-HST project (left panel) and the resulting error in the radio-FIR correlation parameter ($\Delta q_{FIR}$) as a function of spectroscopic redshift (right panel).  The additional scatter in $\Delta q_{FIR}$ resulting from using photometric redshift is 0.112 in dex.
\label{qdiff_photoz}}
\end{figure*}

As discussed in \S~\ref{SPECZ}, we limit our analysis only to the subsample of GN and GS radio sources with a spectroscopic redshift because we aim to remove any additional and possibly systematic noise introduced by adopting photometric redshifts, at the expense of reducing the total sample size by up to 16\%.   As shown in Figure~\ref{qdiff_photoz}, photometric redshifts reported by the 3D-HST project, derived using the well-sampled and deep UV-to-NIR photometry available in these fields, are quite good in general, with a few catastrophic outliers.  When these redshift errors are propagated into the derivation of $q_{FIR}$ as shown on the right panel, the magnitude of additional scatter introduced by using photometric redshifts is 0.112 in dex.  This is about 50\% of the intrinsic scatter measured among the local sample of IR-selected SFGs by \citet{yun01} and thus is substantial in magnitude.  Fortunately, the redshift error and the resulting changes in $\Delta q_{FIR}$ seem random and not systematic, and using photometric redshifts might be acceptable in future studies if the analysis requires a much larger sample size for improved statistics.

\section{B. Catalog of 5~GHz flux densities and spectral indices of our radio sources \label{CAT}}
The final radio source catalog is presented in Table~\ref{tab:catalog}. It includes all 52 GN and 88 GS sources cataloged from images with original beam sizes. The 5~GHz flux densities listed in Table~\ref{tab:catalog} are measured with the original beam sizes, but the spectral index is derived with the beam-matched catalogs as shown in \S~\ref{DATA14}. Eight GS radio sources with original beam sizes are merged into three sources in the image with the beam size matched to that of 1.4~GHz image (refer to \S~\ref{DATA14}). Positions of three merged sources (GS-15, GS-44 and GS-73) are found in the beam matched catalog, but their 5~GHz flux densities are measured from the image with the original beam size. We also list the eight GS sources below the merged sources as GS-15a, -15b, -15c, GS-44a, -44b, -44c, GS-73a, and -73b. The merged sources are not Gaussian-like shapes in the image with the original beam size, so their flux densities are poorly measured by AIPS tasks SAD or JMFIT which utilize the 2D Gaussian fitting function. For this reason, the flux densities of three merged sources are measured with the AIPS task TVSTAT which is appropriate for measuring the flux density of the irregular shaped source. The flux density measured with TVSTAT are larger in general than summation over flux densities of individual sources, because the TVSTAT traces flux densities of regions among individual sources. 

Data columns of Table~\ref{tab:catalog} are summarized as follows: (1) Source ID (ID), (2) Right Ascension (RA J2000), a unit of [hour, minute, second], (3) uncertainty of RA, a unit of second, (4) Declination (DEC J2000), a unit of [$^{\circ}$~ \arcmin~ \arcsec~], (5) uncertainty of DEC, a unit of \arcsec, (6) peak flux density (S$_{peak}$) and its uncertainty, a unit of $\mu$Jy beam$^{-1}$, (7) integrated flux density (S$_{int}$) and its uncertainty, a unit of $\mu$Jy, and (8) radio spectral index ($\alpha$) and its uncertainty.

\begin{center}
\begin{longtable}{rccccccc}
\caption{5~GHz flux densities and spectral indices of GN \& GS radio sources} \label{tab:catalog} \\

\hline {\textbf{ID}} & {\textbf{RA J2000}} & {\textbf{eRA}} & {\textbf{DEC J2000}} & {\textbf{eDEC}} & {\textbf{S$_{peak}$}} & {\textbf{S$_{int}$}}$^{10}$ & {\textbf{$\alpha$}}$^{11}$ \\ 
 & [h m s] & [s] & [$^{\circ}$~\arcmin~\arcsec~] & [\arcsec\,] & [$\mu$Jy beam$^{-1}$~] & [$\mu$Jy] & \\  \hline 
\endfirsthead

\multicolumn{8}{c}%
{{\bfseries \tablename\ \thetable{} -- continued from previous page}} \\
\hline {\textbf{ID}} & {\textbf{RA J2000}} & {\textbf{eRA}} & {\textbf{DEC J2000}} & {\textbf{eDEC}} & {\textbf{S$_{peak}$}} & {\textbf{S$_{int}$}}$^{10}$ & {\textbf{$\alpha$}}$^{11}$ \\ 
 & [h m s] & [s] & [$^{\circ}$~\arcmin~\arcsec~] & [\arcsec\,] & [$\mu$Jy beam$^{-1}$~] & [$\mu$Jy] & \\  \hline 
\endhead

\hline \multicolumn{8}{|r|}{{Continued on next page}} \\ \hline
\endfoot

\hline \hline
\endlastfoot

GS-01 & 3 31 59.619 & 0.034 & -27 47 32.87 & 0.07 & 27.8 $\pm$4.9 & 27.8 $\pm$ 4.9 & 0.265 $\pm$ 0.138 \\ 
GS-02 & 3 31 59.843 & 0.011 & -27 45 40.88 & 0.02 & 96.2 $\pm$ 5.2 & 96.2 $\pm$ 5.2 & 0.727 $\pm$ 0.051 \\ 
GS-03 & 3 32 1.547 & 0.006 & -27 46 47.84 & 0.01 & 550.4 $\pm$ 4.0 & 9338.2 $\pm$ 78.7 & 0.903 $\pm$ 0.001 \\ 
GS-04 & 3 32 3.667 & 0.015 & -27 46 3.98 & 0.03 & 63.8 $\pm$ 4.1 & 66.3 $\pm$ 7.3 & 0.189 $\pm$ 0.061 \\ 
GS-05 & 3 32 6.446 & 0.054 & -27 47 28.96 & 0.08 & 18.2 $\pm$ 3.5 & 25.1 $\pm$ 7.4 & 0.901 $\pm$ 0.083 \\ 
GS-06 & 3 32 8.538 & 0.042 & -27 46 48.55 & 0.06 & 26.7 $\pm$ 3.2 & 55.8 $\pm$ 9.3 & 1.088 $\pm$ 0.044 \\ 
GS-07 & 3 32 8.673 & 0.000 & -27 47 34.68 & 0.00 & 4030.0 $\pm$ 3.0 & 4030.0 $\pm$ 3.0 & -0.521 $\pm$ 0.002 \\ 
GS-08 & 3 32 9.716 & 0.003 & -27 42 48.43 & 0.01 & 329.5 $\pm$ 4.8 & 329.5 $\pm$ 4.8 & -0.168 $\pm$ 0.019 \\ 
GS-09 & 3 32 10.734 & 0.060 & -27 48 7.49 & 0.08 & 19.0 $\pm$ 3.0 & 41.9 $\pm$ 9.0 & 0.408 $\pm$ 0.086 \\ 
GS-10 & 3 32 10.797 & 0.008 & -27 46 28.11 & 0.01 & 92.5 $\pm$ 3.2 & 99.2 $\pm$ 5.8 & 0.518 $\pm$ 0.028 \\ 
GS-11 & 3 32 10.923 & 0.001 & -27 44 15.26 & 0.00 & 1589.5 $\pm$ 4.0 & 1740.9 $\pm$ 7.0 & 0.449 $\pm$ 0.003 \\ 
GS-12 & 3 32 11.501 & 0.017 & -27 48 15.90 & 0.04 & 39.8 $\pm$ 3.1 & 51.4 $\pm$ 6.3 & 0.108 $\pm$ 0.081 \\ 
GS-13 & 3 32 11.532 & 0.014 & -27 47 13.31 & 0.02 & 57.5 $\pm$ 3.1 & 72.9 $\pm$ 6.3 & 0.889 $\pm$ 0.033 \\ 
GS-14 & 3 32 11.615 & 0.048 & -27 50 27.54 & 0.09 & 16.2 $\pm$ 3.2 & 16.2 $\pm$ 3.2 & $<$  0.347 \\ 
GS-15 & 3 32 13.104 & 0.020 & -27 43 50.95 & 0.21 &           & 368.1 $\pm$ 28.5 & 1.312 $\pm$ 0.022 \\ 
15a & 3 32 13.047 & 0.095 & -27 43 50.60 & 0.09 & 25.6 $\pm$ 3.3 & 159.2 $\pm$ 23.2 &  \\ 
15b & 3 32 13.115 & 0.056 & -27 43 51.63 & 0.05 & 32.5 $\pm$ 3.3 & 90.6 $\pm$ 12.2 &  \\ 
15c & 3 32 13.139 & 0.029 & -27 43 50.62 & 0.04 & 42.7 $\pm$ 3.4 & 105.5 $\pm$ 11.1 &  \\ 
GS-16 & 3 32 13.247 & 0.033 & -27 42 41.31 & 0.06 & 30.0 $\pm$ 4.3 & 30.0 $\pm$ 4.3 & 0.751 $\pm$ 0.097 \\ 
GS-17 & 3 32 13.490 & 0.008 & -27 49 53.11 & 0.02 & 87.3 $\pm$ 3.0 & 103.4 $\pm$ 5.9 & -0.604 $\pm$ 0.052 \\ 
GS-18 & 3 32 13.898 & 0.013 & -27 50 0.88 & 0.02 & 56.4 $\pm$ 3.1 & 56.4 $\pm$ 3.1 & $<$ -0.483 \\ 
GS-19 & 3 32 14.164 & 0.051 & -27 49 10.53 & 0.08 & 17.9 $\pm$ 2.9 & 33.8 $\pm$ 7.8 & 0.959 $\pm$ 0.070 \\ 
GS-20 & 3 32 14.213 & 0.053 & -27 46 34.89 & 0.08 & 16.6 $\pm$ 3.0 & 24.4 $\pm$ 6.7 & $<$ 0.338 \\ 
GS-21 & 3 32 14.992 & 0.033 & -27 42 25.49 & 0.07 & 24.8 $\pm$ 4.2 & 24.8 $\pm$ 4.2 & $<$ 0.238 \\ 
GS-22 & 3 32 15.267 & 0.053 & -27 50 19.76 & 0.12 & 15.0 $\pm$ 2.9 & 32.0 $\pm$ 8.5 & $<$ -0.143 \\ 
GS-23 & 3 32 15.338 & 0.043 & -27 50 37.72 & 0.09 & 16.4 $\pm$ 3.0 & 20.9 $\pm$ 6.1 & 0.349 $\pm$ 0.114 \\ 
GS-24 & 3 32 17.157 & 0.019 & -27 43 3.70 & 0.04 & 40.2 $\pm$ 3.6 & 40.2 $\pm$ 3.6 & 0.461 $\pm$ 0.108 \\ 
GS-25 & 3 32 17.183 & 0.032 & -27 52 21.10 & 0.05 & 32.0 $\pm$ 3.3 & 54.1 $\pm$ 8.1 & 0.452 $\pm$ 0.059 \\ 
GS-26 & 3 32 18.023 & 0.002 & -27 47 18.77 & 0.00 & 375.9 $\pm$ 3.0 & 384.9 $\pm$ 5.2 & 0.220 $\pm$ 0.009 \\ 
GS-27 & 3 32 18.563 & 0.044 & -27 51 34.82 & 0.07 & 18.2 $\pm$ 3.1 & 22.6 $\pm$ 6.1 & $<$ 0.048 \\ 
GS-28 & 3 32 19.052 & 0.048 & -27 52 14.99 & 0.09 & 18.2 $\pm$ 3.1 & 32.4 $\pm$ 8.0 & 0.737 $\pm$ 0.115 \\ 
GS-29 & 3 32 19.310 & 0.019 & -27 52 19.52 & 0.04 & 37.7 $\pm$ 3.2 & 44.4 $\pm$ 6.2 & -0.033 $\pm$ 0.103 \\ 
GS-30 & 3 32 19.316 & 0.003 & -27 54 6.58 & 0.00 & 352.4 $\pm$ 4.3 & 2432.7 $\pm$ 60.0 & 0.923 $\pm$ 0.007 \\ 
GS-31 & 3 32 19.514 & 0.012 & -27 52 17.87 & 0.02 & 63.0 $\pm$ 3.2 & 69.3 $\pm$ 6.0 & 0.693 $\pm$ 0.039 \\ 
GS-32 & 3 32 19.817 & 0.012 & -27 41 23.10 & 0.02 & 83.1 $\pm$ 4.6 & 83.1 $\pm$ 4.6 & 0.594 $\pm$ 0.047 \\ 
GS-33 & 3 32 21.285 & 0.016 & -27 44 35.90 & 0.03 & 43.6 $\pm$ 2.9 & 43.6 $\pm$ 2.9 & 1.102 $\pm$ 0.042 \\ 
GS-34 & 3 32 22.159 & 0.058 & -27 49 36.76 & 0.09 & 14.5 $\pm$ 2.9 & 23.3 $\pm$ 6.9 & 0.673 $\pm$ 0.114 \\ 
GS-35 & 3 32 22.281 & 0.032 & -27 48 4.83 & 0.10 & 15.5 $\pm$ 3.0 & 15.5 $\pm$ 3.0 & 0.713 $\pm$ 0.162 \\ 
GS-36 & 3 32 22.514 & 0.017 & -27 48 4.99 & 0.03 & 38.0 $\pm$ 3.0 & 38.0 $\pm$ 3.0 & 0.343 $\pm$ 0.095 \\ 
GS-37 & 3 32 22.597 & 0.028 & -27 44 26.11 & 0.04 & 30.3 $\pm$ 2.9 & 41.5 $\pm$ 6.1 & 0.809 $\pm$ 0.056 \\ 
GS-38 & 3 32 22.723 & 0.037 & -27 41 26.79 & 0.07 & 28.5 $\pm$ 4.1 & 44.8 $\pm$ 9.7 & 0.095 $\pm$ 0.112 \\ 
GS-39 & 3 32 24.262 & 0.039 & -27 41 26.81 & 0.06 & 31.9 $\pm$ 4.0 & 47.9 $\pm$ 9.1 & $<$ -0.859 \\ 
GS-40 & 3 32 24.670 & 0.045 & -27 53 34.37 & 0.09 & 19.5 $\pm$ 3.5 & 24.4 $\pm$ 7.1 & 0.895 $\pm$ 0.108 \\ 
GS-41 & 3 32 25.174 & 0.051 & -27 54 50.31 & 0.09 & 24.1 $\pm$ 4.6 & 30.1 $\pm$ 9.1 & 0.795 $\pm$ 0.086 \\ 
GS-42 & 3 32 25.180 & 0.035 & -27 42 19.15 & 0.06 & 23.1 $\pm$ 3.4 & 27.3 $\pm$ 6.6 & 1.347 $\pm$ 0.193 \\ 
GS-43 & 3 32 26.769 & 0.037 & -27 41 45.98 & 0.08 & 23.9 $\pm$ 3.6 & 36.9 $\pm$ 8.4 & $<$ 0.084 \\ 
GS-44 & 3 32 26.974 & 0.001 & -27 41 7.16 & 0.01 &            & 5390.7 $\pm$ 33.0 & 0.958 $\pm$ 0.002 \\ 
44a & 3 32 26.953 & 0.001 & -27 41 7.88 & 0.00 & 1069.0 $\pm$ 4.0 & 3613.0 $\pm$ 17.0 &  \\ 
44b & 3 32 27.011 & 0.001 & -27 41 5.44 & 0.00 & 1079.0 $\pm$ 4.0 & 1290.0 $\pm$ 8.0 &  \\ 
44c & 3 32 27.060 & 0.044 & -27 41 3.69 & 0.03 & 77.6 $\pm$ 3.9 & 463.6 $\pm$ 27.2 &  \\ 
GS-45 & 3 32 27.018 & 0.072 & -27 42 18.66 & 0.14 & 16.4 $\pm$ 3.2 & 30.1 $\pm$ 8.6 & $<$ -0.020 \\ 
GS-46 & 3 32 27.728 & 0.031 & -27 50 41.24 & 0.05 & 18.9 $\pm$ 2.9 & 18.9 $\pm$ 2.9 & 1.311 $\pm$ 0.177 \\ 
GS-47 & 3 32 28.002 & 0.024 & -27 46 39.65 & 0.04 & 30.0 $\pm$ 2.9 & 39.5 $\pm$ 6.1 & 0.592 $\pm$ 0.060 \\ 
GS-48 & 3 32 28.425 & 0.037 & -27 43 44.85 & 0.08 & 15.1 $\pm$ 2.9 & 15.1 $\pm$ 2.9 & $<$ 0.740 \\ 
GS-49 & 3 32 28.513 & 0.030 & -27 46 58.48 & 0.06 & 22.9 $\pm$ 3.0 & 22.9 $\pm$ 3.0 & 0.864 $\pm$ 0.098 \\ 
GS-50 & 3 32 28.742 & 0.008 & -27 46 20.60 & 0.01 & 94.7 $\pm$ 2.9 & 127.8 $\pm$ 6.1 & 0.534 $\pm$ 0.022 \\ 
GS-51 & 3 32 28.826 & 0.005 & -27 43 55.94 & 0.01 & 127.5 $\pm$ 2.8 & 244.2 $\pm$ 8.8 & 1.554 $\pm$ 0.027 \\ 
GS-52 & 3 32 28.886 & 0.026 & -27 41 29.76 & 0.04 & 38.6 $\pm$ 3.9 & 38.6 $\pm$ 3.9 & $<$ -0.464 \\ 
GS-53 & 3 32 29.876 & 0.036 & -27 44 25.26 & 0.14 & 28.2 $\pm$ 2.5 & 226.1 $\pm$ 22.7 & 1.099 $\pm$ 0.025 \\ 
GS-54 & 3 32 29.986 & 0.101 & -27 44 5.39 & 0.14 & 15.6 $\pm$ 2.6 & 71.7 $\pm$ 14.2 & 1.140 $\pm$ 0.056 \\ 
GS-55 & 3 32 31.489 & 0.055 & -27 46 23.51 & 0.09 & 15.4 $\pm$ 2.8 & 27.4 $\pm$ 7.3 & 1.067 $\pm$ 0.082 \\ 
GS-56 & 3 32 31.546 & 0.008 & -27 50 29.00 & 0.01 & 89.8 $\pm$ 2.9 & 110.9 $\pm$ 5.8 & -0.578 $\pm$ 0.075 \\ 
GS-57 & 3 32 33.007 & 0.033 & -27 46 6.64 & 0.07 & 16.1 $\pm$ 2.9 & 16.1 $\pm$ 2.9 & $<$ 0.597 \\ 
GS-58 & 3 32 33.446 & 0.057 & -27 52 28.55 & 0.07 & 19.0 $\pm$ 2.9 & 38.4 $\pm$ 8.3 & 0.981 $\pm$ 0.062 \\ 
GS-59 & 3 32 36.185 & 0.053 & -27 49 32.17 & 0.08 & 15.1 $\pm$ 2.9 & 20.3 $\pm$ 6.2 & 1.105 $\pm$ 0.107 \\ 
GS-60 & 3 32 37.734 & 0.030 & -27 50 0.71 & 0.05 & 28.3 $\pm$ 2.9 & 38.0 $\pm$ 6.1 & 0.908 $\pm$ 0.084 \\ 
GS-61 & 3 32 37.768 & 0.027 & -27 52 12.63 & 0.05 & 29.5 $\pm$ 3.1 & 36.6 $\pm$ 6.2 & 0.631 $\pm$ 0.061 \\ 
GS-62 & 3 32 37.890 & 0.069 & -27 53 17.86 & 0.15 & 17.1 $\pm$ 3.4 & 30.5 $\pm$ 8.8 & $<$ 0.277 \\ 
GS-63 & 3 32 38.791 & 0.033 & -27 44 49.28 & 0.05 & 22.4 $\pm$ 2.9 & 26.4 $\pm$ 5.5 & 0.633 $\pm$ 0.103 \\ 
GS-64 & 3 32 38.838 & 0.076 & -27 49 56.60 & 0.07 & 15.0 $\pm$ 2.8 & 28.0 $\pm$ 7.5 & 0.136 $\pm$ 0.093 \\ 
GS-65 & 3 32 39.193 & 0.053 & -27 53 57.94 & 0.10 & 22.4 $\pm$ 3.8 & 48.5 $\pm$ 11.5 & 0.384 $\pm$ 0.081 \\ 
GS-66 & 3 32 39.488 & 0.024 & -27 53 1.87 & 0.04 & 40.7 $\pm$ 3.4 & 62.1 $\pm$ 7.8 & 0.607 $\pm$ 0.049 \\ 
GS-67 & 3 32 43.320 & 0.034 & -27 46 47.01 & 0.06 & 19.4 $\pm$ 2.9 & 19.4 $\pm$ 2.9 & $<$ 0.256 \\ 
GS-68 & 3 32 43.542 & 0.045 & -27 54 55.05 & 0.07 & 29.1 $\pm$ 5.8 & 29.1 $\pm$ 5.8 & $<$ 0.271 \\ 
GS-69 & 3 32 44.051 & 0.062 & -27 51 43.90 & 0.19 & 20.9 $\pm$ 2.9 & 105.2 $\pm$ 17.4 & 1.072 $\pm$ 0.039 \\ 
GS-70 & 3 32 44.275 & 0.009 & -27 51 41.31 & 0.02 & 85.1 $\pm$ 3.2 & 106.9 $\pm$ 6.4 & 0.741 $\pm$ 0.024 \\ 
GS-71 & 3 32 45.401 & 0.036 & -27 43 49.36 & 0.08 & 17.2 $\pm$ 3.4 & 17.2 $\pm$ 3.4 & $<$ 0.502 \\ 
GS-72 & 3 32 45.967 & 0.038 & -27 53 16.25 & 0.08 & 25.0 $\pm$ 4.2 & 25.0 $\pm$ 4.2 & 1.641 $\pm$ 0.146 \\ 
GS-73 & 3 32 46.802 & 0.008 & -27 42 14.40 & 0.14 &           & 93.5 $\pm$ 13.0 & -0.265 $\pm$ 0.078 \\ 
73a & 3 32 46.770 & 0.039 & -27 42 12.50 & 0.05 & 34.2 $\pm$ 4.6 & 42.8 $\pm$ 9.2 &  \\ 
73b & 3 32 46.884 & 0.045 & -27 42 15.56 & 0.07 & 29.4 $\pm$ 4.6 & 38.8 $\pm$ 9.4 &  \\ 
GS-74 & 3 32 47.494 & 0.040 & -27 42 43.97 & 0.10 & 21.9 $\pm$ 4.3 & 21.9 $\pm$ 4.3 & $<$ 0.737 \\ 
GS-75 & 3 32 47.902 & 0.047 & -27 42 33.12 & 0.10 & 24.1 $\pm$ 4.3 & 45.2 $\pm$ 11.5 & 1.155 $\pm$ 0.074 \\ 
GS-76 & 3 32 48.185 & 0.031 & -27 52 57.02 & 0.06 & 31.7 $\pm$ 4.1 & 37.7 $\pm$ 8.0 & 0.066 $\pm$ 0.120 \\ 
GS-77 & 3 32 48.566 & 0.040 & -27 49 34.63 & 0.05 & 24.8 $\pm$ 3.0 & 39.4 $\pm$ 7.2 & 0.636 $\pm$ 0.086 \\ 
GS-78 & 3 32 49.440 & 0.002 & -27 42 35.54 & 0.00 & 599.6 $\pm$ 4.7 & 716.9 $\pm$ 9.1 & 1.159 $\pm$ 0.008 \\ 
GS-79 & 3 32 51.838 & 0.020 & -27 44 37.09 & 0.03 & 53.7 $\pm$ 3.7 & 72.2 $\pm$ 7.7 & 0.218 $\pm$ 0.059 \\ 
GS-80 & 3 32 52.077 & 0.008 & -27 44 25.57 & 0.01 & 151.8 $\pm$ 3.8 & 214.6 $\pm$ 8.2 & -0.279 $\pm$ 0.030 \\ 
GS-81 & 3 32 52.326 & 0.055 & -27 45 42.24 & 0.07 & 19.0 $\pm$ 3.4 & 26.1 $\pm$ 7.3 & 0.445 $\pm$ 0.133 \\ 
GS-82 & 3 32 53.863 & 0.045 & -27 51 36.91 & 0.10 & 21.4 $\pm$ 4.1 & 29.3 $\pm$ 8.6 & $<$ -0.035 \\ 
GS-83 & 3 32 59.386 & 0.050 & -27 47 58.50 & 0.08 & 22.7 $\pm$ 4.4 & 28.8 $\pm$ 8.8 & $<$ 0.040 \\ 
\hline
GN-01 & 12 36 0.117 & 0.144 & 62 10 46.92 & 0.16 & 29.0 $\pm$ 5.4 & 46.1 $\pm$ 13.0 & 0.796 $\pm$ 0.101 \\ 
GN-02 & 12 36 1.803 & 0.111 & 62 11 26.34 & 0.12 & 32.7 $\pm$ 5.4 & 32.7 $\pm$ 5.4 & 1.034 $\pm$ 0.064 \\ 
GN-03 & 12 36 3.238 & 0.070 & 62 11 10.67 & 0.07 & 43.9 $\pm$ 5.2 & 43.9 $\pm$ 5.2 & 1.042 $\pm$ 0.049 \\ 
GN-04 & 12 36 6.607 & 0.054 & 62 9 50.91 & 0.06 & 63.0 $\pm$ 4.7 & 90.8 $\pm$ 10.6 & 0.665 $\pm$ 0.044 \\ 
GN-05 & 12 36 8.122 & 0.018 & 62 10 35.70 & 0.02 & 158.2 $\pm$ 4.5 & 169.6 $\pm$ 8.2 & 0.205 $\pm$ 0.018 \\ 
GN-06 & 12 36 8.790 & 0.295 & 62 11 43.57 & 0.15 & 21.6 $\pm$ 4.2 & 60.7 $\pm$ 15.6 & -0.149 $\pm$ 0.098 \\ 
GN-07 & 12 36 12.513 & 0.158 & 62 11 40.22 & 0.16 & 21.4 $\pm$ 4.0 & 39.3 $\pm$ 10.7 & 0.626 $\pm$ 0.099 \\ 
GN-08 & 12 36 17.096 & 0.068 & 62 10 11.35 & 0.06 & 38.0 $\pm$ 3.9 & 38.0 $\pm$ 3.9 & 0.222 $\pm$ 0.052 \\ 
GN-09 & 12 36 19.453 & 0.078 & 62 12 52.47 & 0.09 & 31.9 $\pm$ 4.1 & 31.9 $\pm$ 4.1 & 0.930 $\pm$ 0.054 \\ 
GN-10 & 12 36 20.284 & 0.022 & 62 8 44.12 & 0.02 & 122.9 $\pm$ 4.3 & 133.7 $\pm$ 7.9 & -0.054 $\pm$ 0.023 \\ 
GN-11 & 12 36 21.217 & 0.122 & 62 11 8.68 & 0.17 & 18.2 $\pm$ 3.5 & 25.8 $\pm$ 7.8 & 0.865 $\pm$ 0.112 \\ 
GN-12 & 12 36 22.536 & 0.012 & 62 6 53.70 & 0.01 & 325.8 $\pm$ 6.4 & 325.8 $\pm$ 6.4 & -0.158 $\pm$ 0.008 \\ 
GN-13 & 12 36 31.266 & 0.038 & 62 9 57.66 & 0.04 & 56.5 $\pm$ 3.5 & 56.5 $\pm$ 3.5 & 0.806 $\pm$ 0.028 \\ 
GN-14 & 12 36 32.480 & 0.063 & 62 11 5.19 & 0.07 & 30.2 $\pm$ 3.4 & 30.2 $\pm$ 3.4 & 0.100 $\pm$ 0.073 \\ 
GN-15 & 12 36 34.456 & 0.043 & 62 12 13.01 & 0.05 & 55.8 $\pm$ 3.3 & 85.0 $\pm$ 7.6 & 0.761 $\pm$ 0.036 \\ 
GN-16 & 12 36 34.505 & 0.040 & 62 12 41.00 & 0.04 & 59.8 $\pm$ 3.4 & 78.1 $\pm$ 7.1 & 0.726 $\pm$ 0.036 \\ 
GN-17 & 12 36 35.608 & 0.115 & 62 14 23.97 & 0.14 & 23.0 $\pm$ 3.9 & 33.0 $\pm$ 8.7 & 0.718 $\pm$ 0.104 \\ 
GN-18 & 12 36 37.042 & 0.074 & 62 8 52.16 & 0.09 & 31.3 $\pm$ 4.0 & 31.3 $\pm$ 4.0 & 0.946 $\pm$ 0.055 \\ 
GN-19 & 12 36 40.742 & 0.100 & 62 10 11.33 & 0.18 & 21.9 $\pm$ 3.4 & 44.1 $\pm$ 9.5 & 0.065 $\pm$ 0.116 \\ 
GN-20 & 12 36 41.563 & 0.077 & 62 9 48.16 & 0.08 & 29.7 $\pm$ 3.7 & 29.7 $\pm$ 3.7 & 0.967 $\pm$ 0.052 \\ 
GN-21 & 12 36 42.093 & 0.016 & 62 13 31.29 & 0.02 & 137.8 $\pm$ 3.5 & 147.3 $\pm$ 6.3 & 0.980 $\pm$ 0.020 \\ 
GN-22 & 12 36 42.187 & 0.057 & 62 15 45.22 & 0.07 & 46.3 $\pm$ 4.3 & 54.5 $\pm$ 8.4 & 1.018 $\pm$ 0.058 \\ 
GN-23 & 12 36 44.390 & 0.003 & 62 11 33.05 & 0.00 & 641.0 $\pm$ 3.4 & 963.0 $\pm$ 6.3 & 0.471 $\pm$ 0.018 \\ 
GN-24 & 12 36 46.074 & 0.100 & 62 14 48.58 & 0.09 & 28.3 $\pm$ 3.6 & 42.6 $\pm$ 8.3 & 0.726 $\pm$ 0.072 \\ 
GN-25 & 12 36 46.331 & 0.012 & 62 14 4.58 & 0.01 & 177.7 $\pm$ 3.5 & 177.7 $\pm$ 3.5 & 0.380 $\pm$ 0.014 \\ 
GN-26 & 12 36 46.334 & 0.082 & 62 16 29.25 & 0.08 & 47.2 $\pm$ 4.3 & 95.9 $\pm$ 12.4 & 1.196 $\pm$ 0.046 \\ 
GN-27 & 12 36 46.660 & 0.104 & 62 8 33.15 & 0.09 & 33.2 $\pm$ 4.6 & 41.7 $\pm$ 9.2 & 0.710 $\pm$ 0.083 \\ 
GN-28 & 12 36 49.663 & 0.027 & 62 7 37.97 & 0.03 & 130.6 $\pm$ 5.9 & 130.6 $\pm$ 5.9 & 0.723 $\pm$ 0.021 \\ 
GN-29 & 12 36 50.181 & 0.190 & 62 8 44.80 & 0.22 & 22.0 $\pm$ 4.4 & 59.6 $\pm$ 15.6 & 0.289 $\pm$ 0.092 \\ 
GN-30 & 12 36 51.091 & 0.082 & 62 10 30.91 & 0.08 & 32.3 $\pm$ 3.7 & 45.0 $\pm$ 8.0 & 0.568 $\pm$ 0.067 \\ 
GN-31 & 12 36 51.721 & 0.078 & 62 12 21.36 & 0.08 & 22.6 $\pm$ 3.4 & 22.6 $\pm$ 3.4 & 0.910 $\pm$ 0.066 \\ 
GN-32 & 12 36 52.814 & 0.088 & 62 18 7.95 & 0.10 & 44.9 $\pm$ 5.6 & 66.9 $\pm$ 12.7 & 0.670 $\pm$ 0.070 \\ 
GN-33 & 12 36 52.888 & 0.012 & 62 14 43.97 & 0.01 & 188.1 $\pm$ 3.5 & 205.8 $\pm$ 6.4 & 0.028 $\pm$ 0.018 \\ 
GN-34 & 12 36 53.372 & 0.089 & 62 11 39.33 & 0.16 & 19.7 $\pm$ 3.5 & 23.3 $\pm$ 6.8 & 0.806 $\pm$ 0.109 \\ 
GN-35 & 12 36 55.800 & 0.111 & 62 9 17.32 & 0.11 & 30.4 $\pm$ 4.6 & 45.0 $\pm$ 10.4 & 0.375 $\pm$ 0.087 \\ 
GN-36 & 12 36 59.317 & 0.003 & 62 18 32.46 & 0.00 & 1106.0 $\pm$ 6.0 & 1122.0 $\pm$ 10.0 & 1.202 $\pm$ 0.012 \\ 
GN-37 & 12 36 59.926 & 0.110 & 62 14 49.80 & 0.15 & 18.4 $\pm$ 3.4 & 18.4 $\pm$ 3.4 & 0.316 $\pm$ 0.117 \\ 
GN-38 & 12 37 0.260 & 0.030 & 62 9 9.76 & 0.03 & 114.2 $\pm$ 5.3 & 119.7 $\pm$ 9.5 & 0.766 $\pm$ 0.032 \\ 
GN-39 & 12 37 1.558 & 0.090 & 62 11 46.40 & 0.12 & 28.3 $\pm$ 3.6 & 47.4 $\pm$ 9.0 & 0.593 $\pm$ 0.071 \\ 
GN-40 & 12 37 2.106 & 0.115 & 62 17 34.32 & 0.16 & 26.7 $\pm$ 4.5 & 46.4 $\pm$ 11.4 & -0.286 $\pm$ 0.091 \\ 
GN-41 & 12 37 8.211 & 0.128 & 62 16 59.05 & 0.13 & 21.6 $\pm$ 4.1 & 21.6 $\pm$ 4.1 & 0.514 $\pm$ 0.129 \\ 
GN-42 & 12 37 8.287 & 0.144 & 62 10 56.17 & 0.18 & 23.4 $\pm$ 4.4 & 43.0 $\pm$ 11.7 & 0.348 $\pm$ 0.098 \\ 
GN-43 & 12 37 11.327 & 0.106 & 62 13 30.91 & 0.07 & 30.5 $\pm$ 3.5 & 46.6 $\pm$ 8.1 & 0.769 $\pm$ 0.067 \\ 
GN-44 & 12 37 13.854 & 0.011 & 62 18 26.27 & 0.01 & 321.0 $\pm$ 5.8 & 321.0 $\pm$ 5.8 & 0.564 $\pm$ 0.013 \\ 
GN-45 & 12 37 16.375 & 0.015 & 62 15 12.32 & 0.01 & 153.0 $\pm$ 3.7 & 153.0 $\pm$ 3.7 & 0.126 $\pm$ 0.016 \\ 
GN-46 & 12 37 16.672 & 0.027 & 62 17 33.39 & 0.03 & 108.3 $\pm$ 4.8 & 118.4 $\pm$ 8.8 & 0.869 $\pm$ 0.030 \\ 
GN-47 & 12 37 21.271 & 0.008 & 62 11 29.91 & 0.01 & 416.1 $\pm$ 5.3 & 429.3 $\pm$ 9.4 & -0.129 $\pm$ 0.015 \\ 
GN-48 & 12 37 25.962 & 0.024 & 62 11 28.59 & 0.01 & 314.8 $\pm$ 5.6 & 1174.7 $\pm$ 26.8 & 1.270 $\pm$ 0.014 \\ 
GN-49 & 12 37 30.818 & 0.066 & 62 12 58.75 & 0.07 & 43.1 $\pm$ 5.2 & 43.1 $\pm$ 5.2 & 0.924 $\pm$ 0.050 \\ 
GN-50 & 12 37 34.503 & 0.173 & 62 17 23.45 & 0.14 & 32.3 $\pm$ 6.2 & 55.8 $\pm$ 15.6 & 0.442 $\pm$ 0.102 \\ 
GN-51 & 12 37 36.922 & 0.092 & 62 14 29.51 & 0.13 & 28.4 $\pm$ 5.4 & 28.4 $\pm$ 5.4 & 0.652 $\pm$ 0.076 \\ 
GN-52 & 12 37 42.331 & 0.091 & 62 15 18.19 & 0.11 & 46.6 $\pm$ 6.4 & 62.1 $\pm$ 13.4 & 0.397 $\pm$ 0.084 \\ 
\end{longtable}

\begin{tablenote}
 \item The integrated flux density is the same as the peak flux density for a point source. 
 \item $^{13}$The spectral index $\alpha$ is estimated between 1.4 and 5~GHz using 1.4~GHz images (\citealt{owen18} for the GN and \citealt{miller13} for the GS fields) and 5~GHz images with same beam sizes as those of 1.4~GHz images.
\end{tablenote}
\end{center}

\clearpage

\end{document}